\documentclass{emulateapj}

\input psfig.sty
\newcommand{\be}{\begin{equation}} \newcommand{\ba}{\begin{eqnarray}}
\newcommand{\ee}{\end{equation}} \newcommand{\ea}{\end{eqnarray}}
\def\etal{{\it et al.\thinspace}}
\def\-{{\em{---}}}
\def \mA {\mbox{${\rm ~m\AA} $} }
\def \rr {\mbox{${\rm RR}$} }
\def \rarb {\mbox{${\rm R_AR_B}$} }
\def \rara {\mbox{${\rm R_AR_A}$} }
\def \dd {\mbox{${\rm DD}$} }
\def \dada {\mbox{${\rm D_AD_A}$} }
\def \dadb {\mbox{${\rm D_AD_B}$} }
\def \dr {\mbox{${\rm DR}$} }
\def \darb {\mbox{${\rm D_AR_B}$} }
\def \dara {\mbox{${\rm D_AR_A}$} }
\def \dbra {\mbox{${\rm D_BR_A}$} }

\def \h         {\hbox{$\, h$} }
\def \hinv      {\hbox{$\, h^{-1}$} }

\def\H7{\mbox {$h_{0.7}$}}
\def\mgII{\mbox {\ion{Mg}{2}~}}
\def\feII{\mbox {\ion{Fe}{2}~}}
\def\cIV{\mbox {\ion{C}{4}}~}
\def\siIV{\mbox {\ion{Si}{4}}~}

\def\IZw18{I~Zw~18}
\def\m82{M82}

\def\h{\mbox {\rm H}}

\def\msun{\mbox {${\rm ~M_\odot}$}}

\def\lya{\mbox {Ly$\alpha$~}}

\def\h0{\mbox {~H$_0$}}
\def\q0{\mbox {~q$_0$}}
\def\o3hb{[OIII]$\lambda5007$~/~H$\beta$~}
\def\O1ha{[OI]$\lambda6300$~/~H$\alpha$~}

\def\s2ha{[SII]$\lambda\lambda6717,31$~/~H$\alpha$~}
\def\2z2{HeII~$\lambda4686$~}
\def\z7{[NII]~$\lambda6583$ }
\def\N2{[NII]~$\lambda6583$~/~H$\alpha$~}
\def\16z2{[SII]~$\lambda\lambda6717, 6731$ }

\def\asec{\ifmmode {'' }\else $''~$\fi}  
\def\amin{\ifmmode {' }\else $'~$\fi}    
\def\arcsper{\ifmmode \rlap.{'' }\else $\rlap{.}'' $\fi} 
\def\arcmper{\ifmmode \rlap.{' }\else $\rlap{.}' $\fi} 
\def\sles{\lower2pt\hbox{$\buildrel {\scriptstyle <}
   \over {\scriptstyle\sim}$}} 
\def\sgreat{\lower2pt\hbox{$\buildrel {\scriptstyle >}
    \over {\scriptstyle\sim}$}} 
\def\kms{~km~s$^{-1}$~}

\def\cm3{~cm$^{-3}$}
\def\col{\mbox {~cm$^{-2}$}}
\def\mpc3{~Mpc$^{3}$}
\def\mpc-3{~Mpc$^{-3}$}

\def\fig{{Figure}}

\def\et{{\rm et\thinspace al.}\ }   

\def\apj{ApJ}
\def\apjs{ApJS}
\def\pasp{PASP}
\def\aj{AJ}
\def\mn{MNRAS}

\def\aa{A\&A}

\slugcomment{Preprint: July 14, 2010}
\shorttitle{Transverse Sizes of \cIV\ Absorption Systems Measured
from Multiple QSO Sightlines}
\shortauthors{Martin \et}

\begin{document}

\title{The Size and Origin of Metal-Enriched Regions in the Intergalactic Medium from
Spectra of Binary Quasars}


\author{Crystal L. Martin\altaffilmark{1,2}, 
Evan Scannapieco\altaffilmark{3},
Sara L. Ellison\altaffilmark{4},
Joeseph F. Hennawi\altaffilmark{5,6,7}, 
S. G. Djorgovski\altaffilmark{8},
Amanda P. Fournier\altaffilmark{2}
}

\altaffiltext{1}{Packard Fellow}
\altaffiltext{2}{Department of Physics,University of California, 
Santa Barbara, CA, 93106, cmartin@physics.ucsb.edu}
\altaffiltext{3}{School of Earth \& Space Exploration, Arizona State University, 
P.O. Box 871404, Tempe, AZ, 85287-1404}
\altaffiltext{4}{Department of Physics and Astronomy, 
University of Victoria, Victoria, British Columbia, V8P 1A1, Canada}
\altaffiltext{5}{Department of Astronomy, 601 Campbell Hall, 
University of California, Berkeley, CA 94720-3411, USA}
\altaffiltext{6}{Max-Planck-Institut f¨ur Astronomie, K¨onigstuhl 17, D-69117
Heidelberg, Germany}
\altaffiltext{7}{NSF Astronomy and Astrophysics Postdoctoral Fellow}
\altaffiltext{8}{Division of Physics, Mathematics, and Astronomy,  
Caltech, Pasadena, CA 91125}

\begin{abstract}
We present tomography of the circum-galactic metal distribution
at redshift 1.7 to 4.5 derived from echellete spectroscopy of
binary quasars. We find \cIV systems at similar redshifts in 
paired sightlines more often than expected for sightline-independent 
redshifts. As the separation of the sightlines increases from 36 kpc
to 907 kpc, the amplitude of this clustering decreases. At the largest 
separations, the \cIV systems  cluster similar to  Lyman-break galaxies 
(Adelberger \et 2005a). The \cIV systems are significantly
less correlated than these galaxies, however, at separations less than 
$ R_1 \simeq 0.42 \pm 0.15 $ \hinv comoving Mpc. Measured in real 
space, i.e., transverse to the sightlines, this length scale 
is significantly smaller than the break scale
estimated from the line-of-sight correlation function
in redshift space (Scannapieco \et 2006a). 
Using a simple model, we interpret the new real-space measurement as an
indication of the typical physical size of enriched regions. We
adopt this size for enriched regions and fit
the redshift-space distortion in the line-of-sight correlation function.
The fitted velocity kick is consistent with the peculiar velocity of
galaxies as determined by the underlying mass distribution and 
places an upper limit on the {\it average} outflow (or inflow) speed of
metals. The implied time scale for dispersing metals is larger than the typical 
stellar ages of Lyman-break galaxies (Shapley \et 2001), and we argue
that enrichment by galaxies at $z \ge 4.3$ played a greater role in dispersing 
metals. To further constrain the growth of enriched regions, we discuss 
empirical constraints on the evolution of the \cIV correlation function 
with cosmic time. This study demonstrates the potential of tomography for 
measuring the metal enrichment history of the circum-galactic medium.
\end{abstract}


\section{INTRODUCTION}

In essentially every variant of Big Bang nucleosynthesis, the primordial
universe is composed mostly of hydrogen and helium, with trace
amounts of the light elements Li, Be, and B. Heavier elements,
the {\it metals},  are produced later by stellar nucleosynthesis, and
their mass traces the quantity of stars formed. Since it seems reasonable 
to expect star formation to be confined to galaxies, the discovery of
highly ionized carbon (and later highly ionized Si and O) in the tenuous 
intergalactic hydrogen gas between galaxies came as a considerable surprise
(e.g., Meyer \& York 1987; Lu  1991; Songaila \& Cowie 1996). Since these 
early observations, it has become well established that some enrichment 
persists to low \lya\ column densities, suggesting at least some filamentary 
intergalactic structures have been enriched (Ellison et al. 2000; Schaye 
et al. 2003). These metals are thought to be ejected from galaxies or 
protogalaxies by powerful galactic winds. These winds circulate metals 
through a {\it circum-galactic} medium (CGM). The formation and evolution
of this circum-galactic medium have not been directly measured. 

The history of circum-galactic metal enrichment directly impacts the 
metallicity of the gas accreted by all galaxies, the properties of 
the first generation of stars, and the galactic mass-metallicity relation.
Suggestions for when this enrichment took place span a broad range from
before the formation of most galaxies by explosions of an early generation of 
very massive stars (Madau, Ferrara, \& Rees 2001; Qian, Sargent \& Wasserburg 2002), 
to during or just after reionization at $z \approx 5-10$ by low-mass galaxies (Gnedin \& 
Ostriker 1997; Ferrara \et 2000; Scannapieco, Ferrara, \& Madau 2002), or even 
significantly later, redshifts 2-4 (Adelberger \et 2005b), when the cosmic star formation rate peaked.  
Observations that constrain the period of metal dispersal are therefore of
great interest.

Previously, the evolution of the carbon mass density has provided the
primary empirical constraint on intergalactic enrichment. The carbon 
mass density inferred from the detections of intervening \cIV  
$\lambda \lambda 1548.204, 1550.781$, however,  
depends sensitively on the gas density and the ambient radiation field.
In cosmological models the fraction of carbon in the third ionization 
state reaches its maximum value near $z \approx 5$ (Oppenheimer \et 2009). 
Beyond $z \sgreat\  5$, both
the decreasing ionization fraction {\it and} the decreasing C mass density 
contribute to the decline in the density of \cIV 
systems  (Becker \et 2009; Ryan-Weber \et 2009). At $z < 5$, where 
the density of \cIV systems remains roughly constant (Songaila 2001), 
substantial IGM enrichment may be masked by the decline in the
ionization fraction $n(C^{+3})/n(C)$
with cosmic time (Cooksey \et 2009; D'Odorico \et 2010). 

Compared to mass density measurements, measuring the size of enriched
regions would be less sensitive to the evolution of the gas density and 
ambient radiation field. Size measurements require only metal markers, 
like \cIV  detections, in multiple sightlines. They can directly constrain 
the period of metal dispersal because larger enriched regions at a given 
cosmic epoch require earlier enrichment (Scannapeico 2005). Size measurements 
for \cIV systems are complicated by both the clustering of galaxies and the 
degeneracy between Doppler shifts and line-of-sight distance. By comparing
previous strategies for measuring the size of enriched regions and introducing
a new method in this paper, we aim to establish the viability of size measurements 
and demonstrate their utility for constraining the history of circum-galactic metal 
enrichment.

The physical distance winds travel is thought to be relatively
independent of galaxy mass and redshift (see, for example, \fig~9 of 
Oppenheimer \& Dav\'{e} 2008). The distance at which a wind stalls, and 
material begins to fall back, defines the size of the {\it metal bubbles} 
surrounding individual galaxies.  By circulating metals, galactic winds 
smooth out the heavy elements over the bubble.  Multiple 
sightlines passing through the same bubble will therefore present 
metal-line systems that are correlated. The amplitude of this correlation 
will be roughly constant over all separations smaller than the typical 
size of the bubble. If bubbles blown by different galaxies overlap, then
pairs of sightlines as wide as the scale of the ensemble of coalesced 
bubbles will present a similar degree of correlation. The typical size of an 
{\it enriched region} may therefore reflect either individual bubbles or 
groupings of bubbles.  

{\it The size of the typical enriched region can be directly inferred from 
measurements of the spatial clustering of metal-line systems.} On scales 
larger than the enriched regions, metal-line systems are expected
to cluster like their sources. The correlation amplitude for galaxies depends on
their halo mass {\it and} redshift, i.e., galaxy populations with the same bias
cluster the same way.   Equivalently, the large-scale clustering of \cIV systems 
determines the bias of the galaxy population dominating enrichment (Scannapieco 2005). 
The clustering strength of galaxies rises 
towards smaller scales and steepens on sub-halo scales.\footnote{
     This steepening represents the transition from the two-halo term
     to the one-halo term. It occurs on a somewhat smaller spatial scale than
     the break in the line-of-sight \cIV auto-correlation function.}
At separations smaller than the enriched regions, however, the clustering 
amplitude of metal-line systems will be roughly constant.
The scale at which the \cIV correlation function flattens is therefore directly
related to the typical size of the enriched regions.

Previously, the clustering of \cIV systems could only be measured in redshift-space 
along the line-of-sight.  At large velocity separations, Scannapieco \et (2006a) showed 
that the metal-line systems cluster like Lyman-break galaxies. At redshift 2.9, 
the clustering strength of Lyman-break galaxies decreases with increasing separation 
as $\xi(r) \simeq (r/4.0 \hinv {\rm ~comoving~ Mpc})^{-1.6}$ (Adelberger \et 2005a). 
In contrast to this steep rise, at separations less than 150\kms, the \cIV auto-correlation function 
presents nearly constant amplitude  (Rauch \et 1996; Pichon \et 2003; Scannapieco \et 
2006). This flattening characterizes  measured correlation functions for \mgII and 
\siIV as well (Petitjean \& Bergeron 1990; Scannapieco \et 2006a).  To smooth out the metal-line
absorption over 150\kms in redshift space, enriched regions expanding with the Hubble flow 
must have physical radii of about 500 kpc.\footnote{At $z \simeq 3$, the 
                rate of cosmic expansion was about 300\kms Mpc$^{-1}$.}
The more detailed analysis of
Scannapieco \et (2006a) demonstrated that placing bubbles of metals $\approx 480$ 
physical kpc in radius around halos of mass $\approx 10^{12}$\msun\ provided a 
reasonable fit to the line-of-sight correlation function. This mass scale
is similar to the halo mass, $\approx 10^{11.2}-10^{11.8}$\msun, estimated 
by Adelberger \et (2005a) from the clustering of  Lyman-break galaxies.

Such a large size for the enriched regions is surprising. First,
it is much larger than the galactic impact parameter at which strong 
\cIV absorption is  detected. The \cIV absorption strength 
declines significantly beyond $b\approx 70-90$~kpc (Steidel \et 2010). Second,
it is much larger than the turnaround radii of galactic winds in cosmological simulations
at $z \sim 3$ (Aguirre \et 2001; Theuns \et 2002; Kawata \& Rauch 2007).
In the cosmological simulations of Oppenheimer \& Dav\'{e} (2008), the bubble
radii range from 60-100 kpc over a wide redshift range; and the largest bubbles
have not formed yet at $z \sim 3$. The analytical model of Furlanetto \& Loeb (2003)
predicts that the wind bubbles reach 100  physical kpc at $z \simeq 3$.
This discrepancy with the
larger size indicted by the line-of-sight clustering of \cIV systems motivates
our measurement of \cIV clustering between sightlines. In real space, unlike redshift 
space, distance and velocity information are not mixed. The line-of-sight correlation function may 
not directly measure physical size if enriched regions have significant velocities with
respect to the cosmic expansion. Plausible sources of these velocity kicks include
the peculiar velocities  of galaxies (Davis \et Peebles 1983), 
galactic outflows (Martin \et 2005; Rupke \et 2005; Weiner \et 2009; Pettini \et 2002; 
Shapley \et 2003; Steidel \et 2010), and cold stream accretion (Dekel \& Birnboim 2008).

In this paper, we provide new insight into the distribution of carbon by comparing the 
redshifts of \cIV systems toward binary quasars. Until recently,  no suitable 
sample of binary quasars had been compiled. Lensed sightlines showed highly correlated
\cIV systems on interstellar scales but do not probe scales larger than 
a few kpc (e.g., Rauch \et 1999, 2001, 2002; Ellison et al. 2004).  At the other extreme, 
little correlation was seen in spectra of wide binaries with physical separations 
$\sgreat\ 1.0$~Mpc (Coppolani \et 2006). Spectra of one binary with an intermediate 
separation of $\approx 50$ kpc presented a number of \cIV systems common to both 
spectra (Crotts \et 1994). Drawing targets from Hennawi (2004) and 
Hennawi \et (2006a, 2006b, 2009), we obtained spectra of binary quasars separated by
5 to 130\asec thereby sampling projected separations from 36 to 907 kpc, i.e., 
a range that spans the various suggested scales for the circulation of metals
by galactic winds. Our observations differ from those analyzed recently by 
Tytler \et (2009) in several important ways. Our sample has a smaller median pair separation
(260 kpc vs about 1 Mpc). Our spectra are more sensitive to weak \cIV systems
due to their higher resolution.
And, we examine the redshift evolution of the clustering properties between $1.7 < z < 4.5$. 
(Our median absorption redshift is 3.0 instead of 2.0.)
The age of the universe at redshift 4.3 is just 1.4~Myr. Significant enrichment and evolution
may occur during the 2.4 Gyr between redshift 4.3 and 1.7. Most
importantly perhaps, we introduce the transverse correlation function
to describe the clustering of \cIV systems. This quantitative description
of the metal distribution provides an empirical target (or test) for cosmological 
models.

The paper is organized as follows. We describe the observations, data quality, and 
sample of \cIV systems in \S~\ref{sec:data}. In \S~\ref{sec:results}, we outline 
several ways to count pairs of \cIV systems, compare these pair counts to those
expected for uncorrelated absorption systems, and explain how the selection function
of the survey was modeled.  Appendix~A derives the expressions for
the error in the correlation function measurements.  We make the critical
comparison between clustering in velocity space and physical space in Section~\ref{sec:discussion}.
We discuss the size of metal-enriched regions and the problems with a previously used maximum likelihood 
estimator. We place an upper limit on the average outflow speed in \S~\ref{sec:discuss_v}.
In  \S~\ref{sec:discuss_origin}, we constrain the in-situ dispersal of metals at $z \simeq 3$ 
and argue that the size of the enriched regions requires significant enrichment by winds at 
an earlier epoch. We propose an evolutionary test in \S~\ref{sec:evolve} that may empirically
pin down the enrichment history. We summarize the conclusions in Section~\ref{sec:summary} discuss.

Atomic data from Morton (2003) and a cosmology with $\Omega_{m} = 0.3$, $\Omega_{\Lambda} = 0.7$,
and $H_0 = 70$ \kms Mpc$^{-1}$ are used unless noted otherwise. 
At $z = 3$, these parameters mean that cosmic expansion produces a velocity shift
of 150\kms across a physical length of 480 physical kpc, which corresponds to
1.35 \hinv comoving Mpc.

\section{OBSERVATIONS} \label{sec:data}


Quasars provide brighter targets for IGM tomography than do galaxies. The
separation of lensed quasars is too small, however, to probe circum-galactic
scales. Finding binary quasars with appropriate separations for CGM studies 
required systematic searches near SDSS and 2dF quasars (Hennawi 2004). 
The SDSS and 2dF quasar samples alone contain few binaries at the relevant
separations (roughly 5 arcseconds up to an arcminute) because the optical
fibers of spectrographs cannot be placed this close. At proper transverse 
separations $l < 1$\hinv Mpc, Hennawi \et (2006a) identified 221 new quasar 
pairs with $0.5 < z_{QSO} < 3.0$. 
Another 27 high-redshift binaries, $2.9 < z < 4.3$, were recently
confirmed to have proper transverse separations from 10~kpc to 650~kpc 
(Hennawi \et 2009). From these samples, we selected bright binary quasars that 
had similar redshifts. Our targets are listed in Table~\ref{tab:esiALL}. 
They form 29 pairs, including one triplet.

At the W. M. Keck Observatory, we obtained spectra of 55 quasars 
with the ESI spectrograph (Sheinis \et 2002). The echellete format provided 
continuous spectral coverage from 4000~\AA\ to 10000~\AA\ at 60\kms (filled slit)
resolution. Exposure times were chosen to reach a S/N ratio of 15 to 20 per 
pixel. {\it This spectral quality provides sensitivity to \cIV components with 
column density  $N(\cIV) \approx 10^{13}$\col. }

\begin{figure}
 \hbox{\includegraphics[scale=0.7,angle=-90,clip=true]{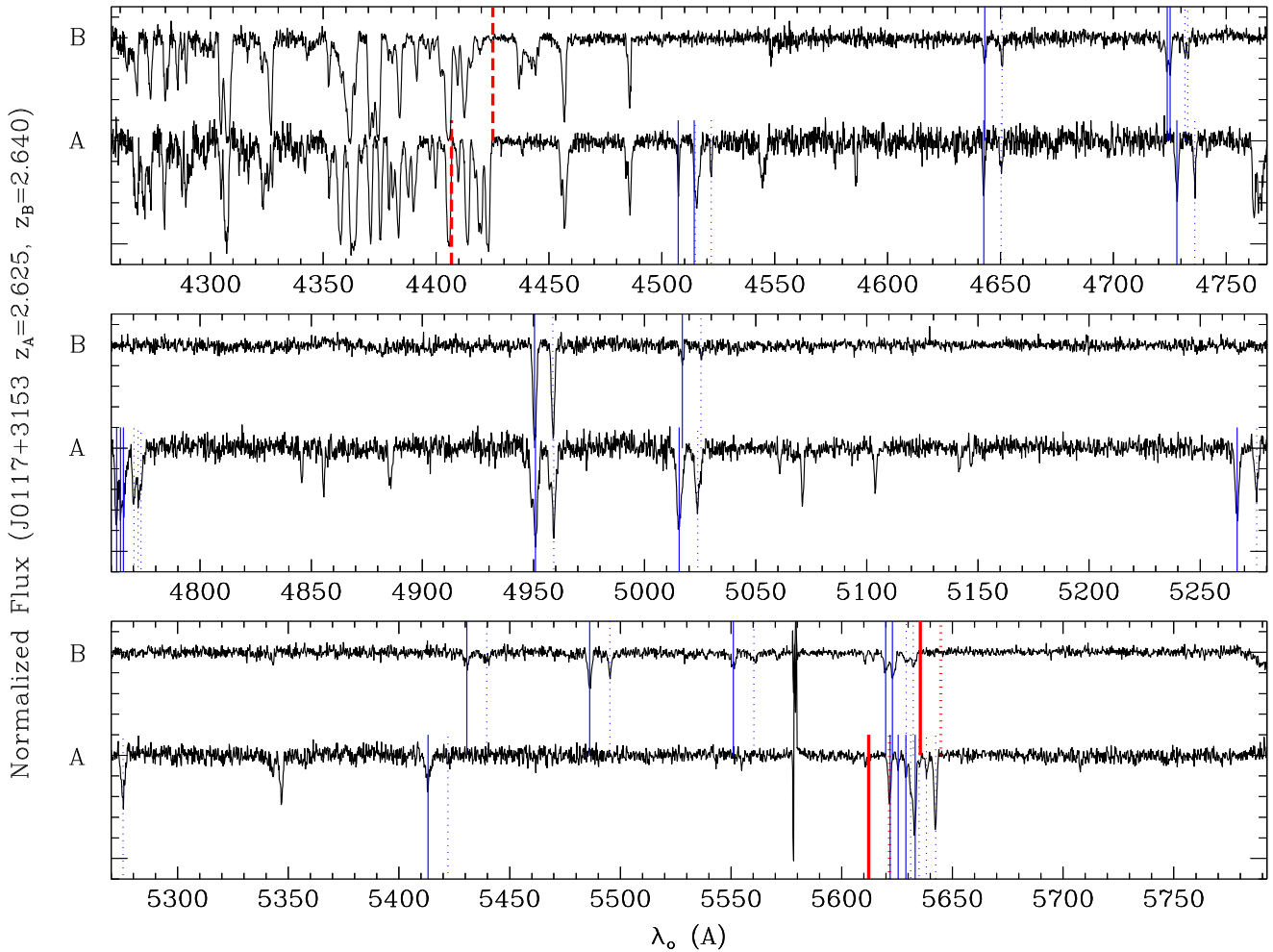}
         \hfill}
 \caption{Example of sightlines separated by 100 physical kpc.
Thick, vertical lines (red) mark \lya and \cIV at the quasar redshift. Thin
vertical lines (blue) mark each \cIV system. We consider all pairs of \cIV
systems between sightline {\it A} and  sightline {\it B}. Provided 
$\Delta v < c$, their velocity separation is 
$\Delta v = c (1 + z_{AB})^{-1} |z_A - z_B| $, where $z_{AB}$ 
is the average redshift of the two systems.  A small velocity 
separation identifies {\it coincident systems}. The number of
coincident system greatly exceeds the number we would find in
the absence of any correlation between the sightlines.
}
 \label{fig:binary_data} \end{figure}

The ESI spectra were reduced using the ESIRedux\footnote
                {http://www.ucolick.org/$\sim$xavier/ESIRedux/index.html}
data reduction pipeline. Scattered light was fitted and subtracted
from the spectroscopic flatfields. One-dimensional spectra were
extracted for each order and averaged using a S/N weight function 
that varied smoothly across the order. Division of each spectral order 
by the fitted quasar continuum left flat, normalized orders that were 
combined into a one-dimensional spectrum. 
Figure~\ref{fig:binary_data} shows example spectra of two quasars
separated by 100 physical kpc.

\subsection{Sample of \cIV Systems}

In each quasar spectrum, we identified \cIV\ absorption lines 
by the doublet spacing. We required the two transitions have 
consistent velocity structure. In a few cases, this revealed
blends between \cIV and unrelated lines. We searched the 
bandpass between the \lya\ forest and the observed wavelength 
of \cIV\ at the quasar redshift. Each pair of spectra probes \cIV 
absorption over a similar redshift interval because the quasars 
were selected to have roughly the same redshift.

For illustration, the 1548 and 1551 
transitions are marked by solid and dotted lines, respectively, in 
the J0117+3153 spectra in Figure~\ref{fig:binary_data}. 
For absorption troughs with well separated 1548 and 1551\AA\ lines,
as seen for example at $z=1.9987, 2.0540, 2.4018, {\rm ~and~} 2.4961$ 
towards J0117+3153-A, we directly integrated the equivalent width in each 
line, $W_{obs} = \int 1 - I(\lambda)/ I_c(\lambda) d\lambda$. We
fitted a pair of Gaussian line profiles to measure the absorber redshift.

When the 1548 and 1551\AA\ lines were blended, we could not directly
integrate their equivalent widths. Multiple velocity components
were typically required to model these broad absorption troughs.
We fitted Gaussian profiles using the {\sc SPECFIT} program.\footnote{
              A  brief  description  is  given  by  G.  Kriss (1994).}
We defined {\it \cIV\ systems} by linking components separated by less than
50~km~s$^{-1}$, the  effective spectral resolution (for the atmospheric seeing).
We set the system redshift equal to the weighted average of the 
component redshifts. The system's equivalent width was calculated
from the formula above using the fitted model of the absorption trough
in place of the data. Linking yielded 316 {\it intervening} 
\cIV\ systems and 134 {\it associated} systems.  Associated systems lie
within 5000\kms of the QSO (in the same sightline) and will not be
included in the sample discussed  in this paper.

\begin{figure}
  \hbox{\includegraphics[scale=0.7,angle=-90,clip=true]{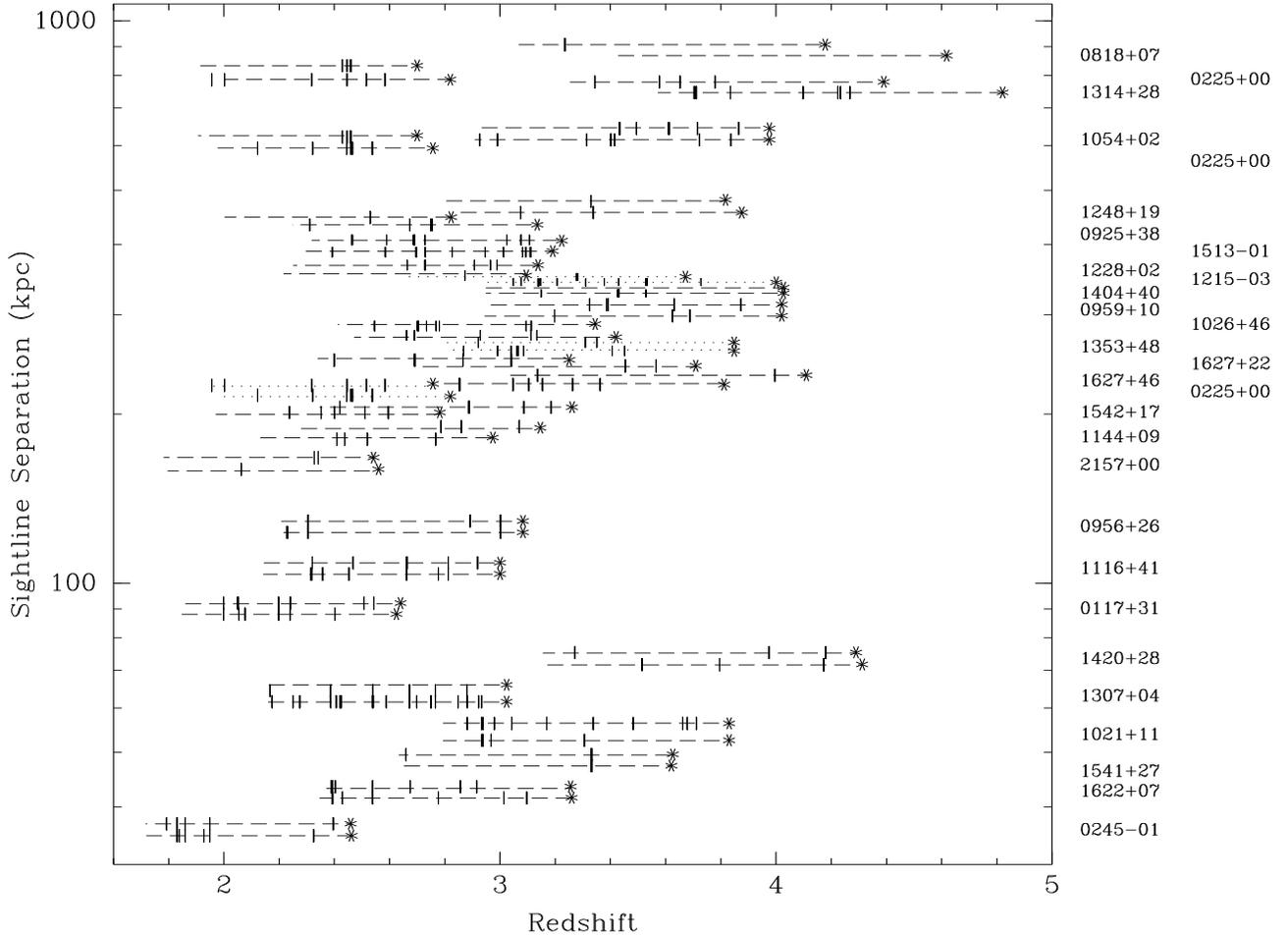}        }
          \caption{Sightlines as a function of binary
            quasar separation and redshift probed.
            Each sightline is shown at the separation of the binary quasar.
            The quasars are marked by asterisks. A vertical tic marks \cIV 
            systems along each sightline.
            (Small vertical offsets have been applied to prevent overlapping
            symbols.)
          }
           \label{fig:sample} \end{figure}

Figure~\ref{fig:sample} shows the redshift of intervening \cIV\ systems
along each sightline. The paired sightlines are ordered by the separation
of the binary quasars. The median sightline separation is 260 kpc with 
a range from 36 kpc to 907 kpc. The absorber redshifts range from 
1.7 to 4.3 with a median of 3.0. The spectra of J1420+28, J1021+11,
J1541+27, and J1622+07 provide critical coverage of small spatial
scales beyond the median redshift.

\subsection{Distribution Function}

As discussed in detail previously (Ellison \et 2000; Songaila 2001; and Scannapieco \et 2006a),
the number of \cIV\ systems in a quasar spectrum increases rapidly with improved sensitivity
to column density. Most of the systems in our sample with rest equivalent width, $W_r(1551)$, 
less than  200~m\AA\ are optically thin. Their measured doublet ratio, $W_{obs}(1548)/ W_{obs}(1551)$, 
is consistent with 2. Their \cIV column density follows directly from their equivalent width, 
$N(\cIV) \approx 10^{14.0}\col (W_{r,1551}/200 {\rm ~m\AA})$. 
This relation provides only a lower limit on the column density of
optically thick systems, however. At the modest dispersion of our spectra, optically thick 
lines need not be black at line center. We identified many optically-thick systems
by their smaller doublet ratio, $ 1 \le W_{obs}(1548)/ W_{obs}(1551) < 2$. We
therefore describe the distribution of line strengths by their equivalent-width distribution
instead of their column density distribution.

Distribution functions for intervening absorbers are usually normalized by redshift 
path rather than redshift. A population with constant comoving space density and 
constant proper size produces an equal number of systems per unit redshift path, $X(z)$, 
along a sightline. For the cosmology adopted here, we have $ dX \equiv (1 + z)^2 [\Omega_{\Lambda} + 
\Omega_m (1+z)^3]^{-0.5}dz$ (Scannapieco \et 2006a).  
We used our sample of intervening \cIV systems to compute the equivalent width
distribution function, $f(W)$, defined as the number of systems found per unit redshift 
path per unit rest equivalent width as follows. 

\begin{figure}
   \hbox {\includegraphics[scale=0.35,angle=-90,clip=true]{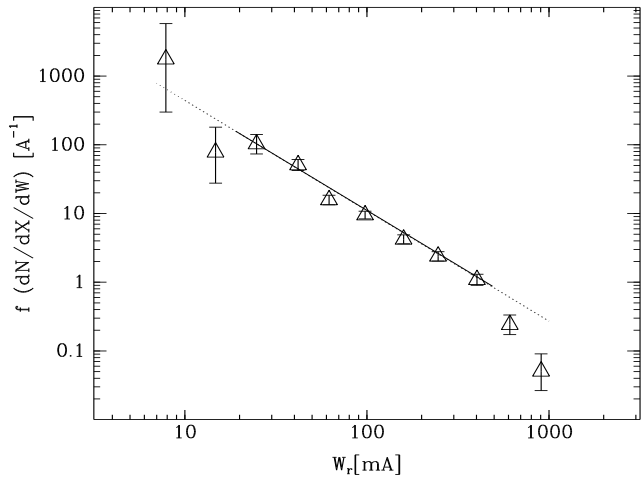}
     \hfill}
      \caption{Equivalent width distribution function for intervening \cIV 1551.
        Components have been grouped into systems using a linking length of 50\kms.
	We fit the distribution function with a power law of the form,
	$ \log f(\AA^{-1}) = b + m \log W(mA)$, including errors in $W$ equal
	to one-half the bin width; see text. 
        The uncertainty $\delta f_k$ follows from the upper and lower 84.1\% confidence level
        Poisson limits, corresponding to $1\sigma$ limit for Gaussian statistics,
        on the number of objects in bin $k$. 
	We find  $b = 4.3 \pm 0.4$ and $m = -1.6 \pm 0.2$;
	or, for $W$ in \AA, $b = -0.6 \pm 0.4$. 
         }
         \label{fig:few} \end{figure}

For each \cIV system, $i$, in our sample, we calculated the total redshift path over which the system
could have been detected, $\Delta X_i = \sum_{j=1}^{N_{QSO}} \Delta X_{i,j}$. 
Detection 
at redshift $z$ requires the observed equivalent width of the weaker \cIV transition exceed 
the $5\sigma$ measurement uncertainty for a system of velocity width 230\kms
at the observed wavelength.\footnote{Spectra containing the 
      wavelength-dependence of the
      continuum S/N ratio for each spectrum are available from the lead author.}  
This width is typical  of  a system near the median equivalent width in our sample.
For each equivalent width bin, we calculate the number of systems per unit redshift path,
\begin{eqnarray}
\left( \frac {d {\sf N}}{dX} \right)_k =   \sum_{i=1}^{{\sf N}_{k}} \frac{1}{\Delta X_i},
\end{eqnarray}
where ${\sf N}_{k}$ is the number of systems in bin $k$.
Normalization by the bin width, 
$\Delta_k \log W_{r,1551} \equiv \log W_{max,k} - \log W_{min,k}$,
yields the distribution function,
\begin{eqnarray}
f(\langle W\rangle_k) = \Delta_k^{-1}\left( \frac {d {\sf N}}{dX} \right)_k.
\end{eqnarray}
at the mean (rest-frame) equivalent width, 
\begin{eqnarray}
\log \langle W\rangle_k \equiv {\sf N}_{k}^{-1} \sum_{i=1}^{{\sf N}_{k}} \log W_i,
\end{eqnarray}
of systems in that bin. Figure~\ref{fig:few} shows the resulting distribution function.
Over nearly two decades in equivalent width, it is well described by a power law. 
We attribute the turnover at large equivalent width,  $W_{r,1551} \sgreat\ 500$~m\AA, to
the non-linear relation between equivalent width and column density. The relatively 
small redshift path observed at the highest sensitivity causes the large error 
bars at $W_{r,1551} < 20$~m\AA. The fit to the seven points between 20 and 500m\AA
is shown in \fig~\ref{fig:few}. This fit to the distribution function
is used in our Monte-Carlo simulations in \S~\ref{sec:results}.
At low equivalent width,  a direct comparison can be made to the column density distribution
functions. We find consistency with Ellison \et (2000), Songaila (2001), and Scannapieco \et (2006a).

\section{CORRELATION FUNCTIONS FOR \cIV SYSTEMS} \label{sec:results}

Two approaches are used here to describe the clustering of \cIV systems.
The first requires multiple sightlines at relatively small separations.
The second, which has been applied previously, estimates separation in 
redshift space instead of real space. The two methods therefore measure
different physical quantities. This section describes the first method
in some detail and applies both approaches to the data. Readers interested 
primarily in the insight gleaned from comparing the two methods may skip
directly to \S~\ref{sec:discussion}.

\subsection{Cross-Correlation of \cIV Systems between Sightlines}


\subsubsection{Pair Counts}

\begin{figure}
 \hbox{
  \includegraphics[scale=0.35,angle=-90,clip=true]{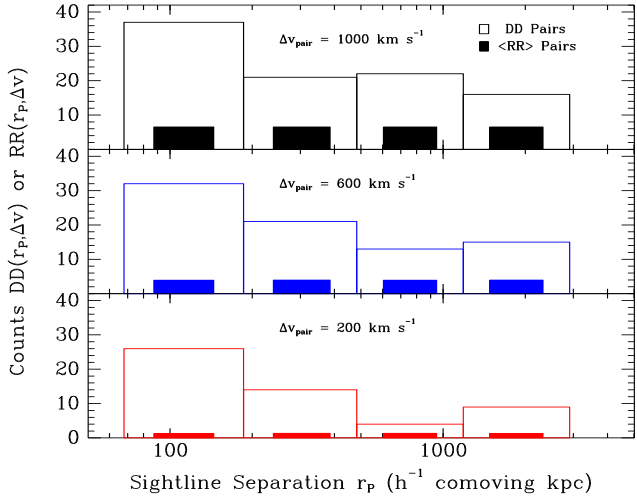}
   \hfill}
       \caption{Number of pairs of \cIV systems as a function of
         sightline separation, $r_P$.
         For each \cIV system in sightline {\it A}, 
         we computed the velocity separation between it and every \cIV system in sightline 
         {\it B}.  Velocity separation smaller than a threshold, $\Delta v_{pair}$,  
	   identify coincident systems -- pairs of \cIV systems at nearly the same redshift 
         in both sightlines.  We sum the number of coincident systems over
	 all the binary quasars $l$ in the bin
         so that  $\dd(r_P,\Delta v_{pair})   \equiv \sum_l {\dd}^l(r_P,\Delta v_{pair})$. 
         {\it The data (open histogram) present many more 
         systems at similar redshift than would
         be expected from uncorrelated sightlines (filled histogram).}
	 The bin boundaries were chosen to have a similar number of 
	 random-random pairs, where the expectation value,
        $\rr(r_P,\Delta v_{pair})  \equiv \sum_l 
	 \langle{\rr}^l(r_{P},\Delta v_{pair})\rangle$.
	 was obtained from a Monte Carlo simulation.
         From bottom to top, the three panels show how the 
	 number of coincident systems declines as the 
	 definition of a pair is relaxed to include larger velocity differences.  
     }
      \label{fig:xi_hist} \end{figure}

The number of absorption-line systems found at similar redshift, but in different 
sightlines, is related to the correlation function on the scale of
the sightlines separation. To find this relationship, {\it data-data} pairs 
can be defined by the apparent velocity difference of the \cIV systems in two 
sightlines. For example, in Figure~\ref{fig:binary_data}, pairs of \cIV systems are 
detected in sightlines A and B at $\lambda $ 4642, $\lambda $ 4728, $\lambda $ 4950, and 
$\lambda $ 5015. Whether or not the system at $\lambda $ 5412 in sightline A and 
the system at $\lambda $ 5430 in 
sightline B count as a pair depends on the velocity threshold used in the definition. In this
example, their velocity separation  is about 950\kms, where $\Delta v = c (1 + z_{AB})^{-1} 
|z_A - z_B| $ and $z_{AB}$ is the average redshift of the two systems. Histograms of
pair counts are illustrated in Figure~\ref{fig:xi_hist} for three values of the threshold 
velocity, $\Delta v_{pair} = 200, 600, {\rm ~and~} 1000$\kms. The counts of systems separated 
by $\Delta v_{AB} < \Delta v_{pair}$ have been summed over all pairs $l$ of sightlines at 
similar transverse separation $r_P$. For the most inclusive definition, all five pairs 
detected towards J0117+3153 contribute to the second bin at 300\hinv comoving Mpc. 
At fixed transverse separation, the number of data-data pairs
declines with more restrictive pair definitions, i.e., smaller $\Delta v_{pair}$. 
For any value of $\Delta v_{pair}$, however, the counts in \fig~\ref{fig:xi_hist} 
increase as the separation of the pairs decreases.

\begin{figure}
 \hbox{\includegraphics[scale=0.35,angle=-90,clip=true]{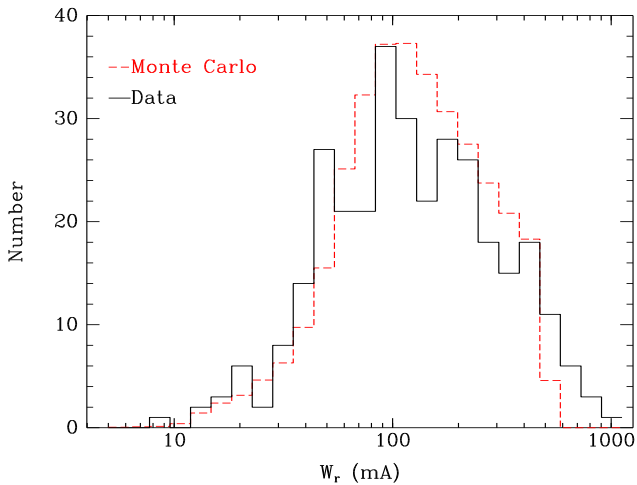}
         \hfill}
    \caption{Distribution of accepted \cIV $W_{r,1551}$ equivalent widths. Dashed, red
histogram compares the fake absorbers to the data, represented by the solid histogram.
The distribution function for fake absorbers was cut-off at 500~m\AA\ in the Monte
Carlo simulation, and the comparison is void for stronger systems. The fake and real data
are consistent within their uncertainties, where the statistical error in the fake
data is $3.1\%$  after 1000 Monte Carlo iterations.
      }
 \label{fig:h2} \end{figure}

We can statistically differentiate physically related pairs of \cIV systems from 
chance coincidences. The number of unrelated pairs, hereafter random-random 
pairs $\rarb(r_P,\Delta v_{pair})$, was modeled using a Monte-Carlo simulation. The velocity 
widths of the fake systems were fixed at 230\kms. Equivalent widths were drawn 
from the distribution function shown in \fig~\ref{fig:few}. The redshifts of 
the fake systems were randomly chosen within the survey range for each quasar
sightline. In the Monte Carlo simulations, fake systems were rejected if the
observed-frame equivalent width was less than the $5\sigma$ detection limit.
Because the variance is a strong function of observed wavelength in each spectrum,
and spectral quality varied by sightline, the redshift distribution of the fake 
systems reflects the sample biases introduced by telluric features, quasar emission 
lines, and intervening absorption lines. The equivalent widths of the fake systems
are shown in Figure~\ref{fig:h2}. Their distribution appears consistent with the real 
sample. Averaged over 1000 iterations of the simulation, the average number of fake 
systems  in a given sightline agrees (to within the Poisson error) with the actual
number of \cIV systems found. When we compare sightlines, these fake systems 
yield far fewer pairs of absorbers.

In each bin of the histogram shown in Figure~\ref{fig:xi_hist}, the ratio of
data-data pairs, $\dd(r_P,\Delta v_{pair})$ to random-random pairs $\rarb(\Delta v_{pair})$  
defines a cross-correlation amplitude, where $\xi_{AB} + 1 = \dadb / \rarb$.  
If $\dadb = \rarb$, then all the pairs are random coincidences; and no clustering is detected.
The pair counts are implicitly a function of the 
binning, so we drop the explicit reference to the binning.
The variance in the ratio $ \dd / \rr $ is generally larger than the Poisson 
error  in the bin counts. The correction to the Poisson term is significant 
when $\dd > N_{abs} / 4$ (Mo \et 1992). In most bins, the number of \dd pairs
is high compared to the number of absorbers, $N_{abs}$, in sightlines contributing
to the bin. Calculating the uncertainty of the measured correlation therefore 
required special attention.

\begin{figure}
 \hbox{
  \includegraphics[scale=0.35,angle=-90,clip=true]{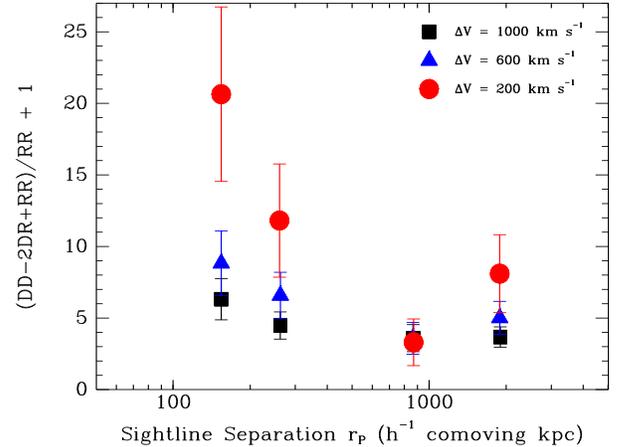}
   \hfill}
       \caption{Cross-correlation amplitude of \cIV systems
         vs. the separation of the sightlines. The value of 
         the Landy \& Szalay (1993) estimator is shown at
         the median separation in each bin.
         The $1\sigma$ error bars include a correction to the 
         Poisson term as described in Appendix~A.
         Decreasing the maximum velocity difference
         defining pairs of \cIV systems, 
         $\Delta v_{pair}$, 
         increases the correlation amplitude between the
         sightlines. For each value of $\Delta v_{pair}$, 
         the correlation amplitude declines as the
         sightline separation increases. 
     }
      \label{fig:xi_dv} \end{figure}

Landy \& Szalay (1993) introduced the estimator, $\xi = (\dd -2\dr + \rr) / 
\rr$, to minimize the variance in a correlation function. For the cross-correlation
of sightlines, the mean number of data-random pairs is
$\langle 2\dr \rangle \equiv \langle \darb \rangle  + \langle \dbra \rangle$, 
where the average is taken over 1000 fake sightlines with the same selection
function as the corresponding quasar spectrum.
The correlation strength, $\xi_{AB}(r_P, \Delta v_{pair})$, calculated from
these pair counts is shown in \fig~\ref{fig:xi_dv}. Since $\xi > 1$ for these data, 
the Landy \& Szalay estimator does not reduce the variance to the Poisson level.
Appendix~A explains how we computed the correction to the Poisson term. These 
larger error estimates are shown for all correlation function figures. 

In \fig~\ref{fig:xi_dv}, the widest sightlines at $r_P = 2.844$ \hinv comoving 
Mpc present highly correlated \cIV systems. The systems cluster even more strongly
as the sightline separation decreases towards $0.075$ \hinv comoving Mpc. The
amplitude of the correlation is independent of the chosen binning in $r_P$.

As $\Delta v_{pair}$ increases, however, the correlation becomes weaker
in \fig~\ref{fig:xi_dv}. Stated another way, the probability that \cIV 
systems (in different sightlines) are physically related declines as
their redshift separation increases. This sensitivity of clustering strength
to $\Delta v_{pair}$ presents an obstacle for comparing the clustering of 
\cIV systems to other populations.

\subsubsection{Transverse Correlation Function} \label{sec:transverse}

One measure of clustering that can be evaluated for our sample,
and compared to other populations, is the projection of the redshift-space 
correlation function.  The projection of $\xi^{\prime}(r)$  along the 
line-of-sight is of particular interest. In redshift space, the 
separation of two systems is $r^2 = r_P^2 + \Delta \Pi^2$.
The distance in redshift-space is calculated 
from the velocity difference of two systems: 
\begin{eqnarray}
\Delta \Pi \equiv (1 + z) \Delta v / H(z) 
\end{eqnarray}
in comoving units. 
The integration limits can be chosen to integrate over peculiar 
velocities, thereby robustly identifying physically related pairs 
of \cIV systems. This projection
\begin{equation}
w(r_P,\Pi_{pair}) \equiv \int_{-\Pi_{pair}} ^{+\Pi_{pair}} \xi^{\prime}(r_P,\Pi) d\Pi,
\end{equation}
defines the transverse correlation function. It differs from the
angular correlation function in that we use the comoving distance
transverse to the sightlines $ r_P(z)$, rather than the 
angle between the sightlines, to describe separation in
the plane of the sky. This approach makes sense because 
$r_P(z)$ is nearly constant over the projection interval of interest.

\begin{figure}
    \hbox{
     \includegraphics[scale=0.35,angle=-90,clip=true]{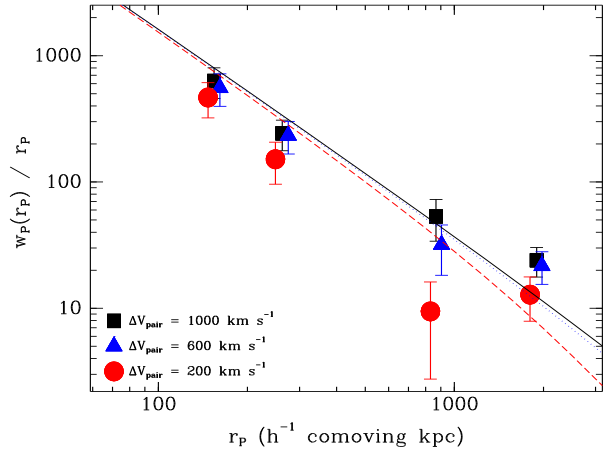}
      \hfill}
       \caption{Transverse \cIV correlation function normalized by sightline 
	 separation versus median sightline separation. The bins are identical 
         to those in Figures~\ref{fig:xi_hist} and \ref{fig:xi_dv}. 
         The lines show the  projection of the galaxy-galaxy 
	 correlation function at $z \simeq 2.9$ (Adelberger \et 2005b),
         which has been extrapolated below 1\hinv comoving Mpc with 
         constant power-law index $\gamma = 1.6$ and perturbed 
	 by linear infall (Hawkins \et 2003). Because infall
	 reduces the volume in redshift space, it increases the correlation 
	 amplitude slightly. Independent of projection interval, however, the 
	 clustering amplitude of  \cIV systems is very similar to that of 
	 Lyman-break galaxies. No parameters are fitted in this figure, so
         their similarity demonstrates that these galaxies and \cIV systems have 
         nearly the same correlation length.
     }
      \label{fig:w_dv_data} \end{figure}

Figure~\ref{fig:w_dv_data} shows the transverse correlation function.
It is the product of $\xi_{AB}(r_P,\Delta v_{pair})$ and the (comoving) distance 
$2 \Delta \Pi_{pair}$. Since the transverse correlation function $w$ has dimensions 
of length, we normalizing it by the sightlne separation and plot the dimensionless 
quantity $w/r_P$.  The \cIV correlation amplitude increases as the sightline 
separation decreases. For the two points at the smallest separations,
the amplitudes of $w/r_P$ are essentially identical for the three projection 
intervals illustrated.  At larger separations the correlation strength presents
a measurable, although still small, dependence on the the projection interval.

To illustrate the influence of the projection interval on $w/r_P$, consider
the projection of Lyman-break galaxies. Their clustering amplitude in 
redshift-space $\xi^{\prime}(r)$ is higher than $\xi(r)$ in real-space
because cosmological infall causes structures to appear smaller in redshift 
space than their true physical size. The reduced volume in redshift-space 
increases the clustering amplitude. To estimate the boost to the galaxy --
galaxy correlation function, we apply Equation~15 of Hawkins \et (2003) which
describes linear infall.\footnote{
        Although linear infall theory is not strictly valid on these scales, 
        Cooray \& Sheth (2002) argue that it is a good approximation. }
The magnitude of the correction depends on the bias. Figure~\ref{fig:w_dv_data}
shows the projection for $b = \sigma_{8,g} / \sigma_{8,CDM} = 2.4$,
$r_0 = 4.0$\hinv comoving Mpc, and $\gamma = 1.6$ (Adelberger \et 2005a). 
At sightline separations $\sgreat\ 1\hinv$, this infall correction noticeably 
boosts the correlation amplitude for each projection interval $\Delta \Pi$; but 
it has no detectable effect at smaller separations. The galaxy-galaxy correlation
function is {\it not} fitted to the \cIV correlation function in Figure~\ref{fig:w_dv_data}. 
It is therefore remarkable that the amount of \cIV clustering agrees roughly with the 
amplitude of the galaxy -- galaxy correlation function. 

Previous work revealed a strong cross-correlation amplitude 
between Lyman-break galaxies and \cIV systems on scales $\sgreat\ 
1$\hinv comoving Mpc (Adelberger \et 2005b). We would like to
explore the relationship of galaxies and \cIV systems on the 
smaller scales probed by the binary quasars, but this turns
out to be difficult with current data. Adelberger \et (2005a) only 
measured the galaxy -- galaxy correlation function at separations 
greater than 1\hinv comoving Mpc. On sub-halo scales, the galaxy -- 
galaxy correlation function is expected to be steeper than it
is in the two-halo regime. By extrapolating the $\gamma = 1.6$
power-law fitted on large scales, the amplitude of the transverse 
correlation function for Lyman-break galaxies is a lower limit
at small separation in Figure~\ref{fig:w_dv_data}. This lower 
limit for galaxies is higher than the upper limit (one standard
deviation) for \cIV systems at small scales. In light of
the uncertain magnitude of this discrepancy, this result alone
is hardly compelling evidence for a difference in the clustering
properties of Lyman-break galaxies and \cIV systems on small
scales. 

We observed only a few binary quasars at separations greater
than 1\hinv comoving Mpc, so our data alone poorly constrain
the power-law index of the correlation function on large
scales. Taking $\gamma = 1.6$ (like LBGs) as a prior, 
the fitted correlation length $r_0 = 3.65  \pm 0.34$\hinv 
comoving Mpc for our sample technically agrees with the value 
$r_0 = 4.0 \pm 0.6$ fitted by Adelberger \et (2005a). The
lower value, however, reflects a best fit with a
lower correlation amplitude  overall. This difference is
worth noting because our sightlines at separations
greater than 1\hinv comoving Mpc show a slightly stronger
correlation in \fig~\ref{fig:w_dv_data} than do galaxies.
There seem to be two options. Our measurement can be fit
with a larger correlation length $r_0 = 7.0 \pm 2.0$\hinv 
comoving Mpc  and shallow power-law index  $\gamma = 1.0 \pm 0.2$.
Or, we can adopt that prior that \cIV systems cluster 
like Lyman-break galaxies on large scales and find
that \cIV do not cluster as strongly as Lyman-break galaxies
on scales $\sles\ 1$\hinv comoving Mpc. To resolve this
ambiguity, we re-examine the clustering of \cIV systems
along the line-of-sight before returning to the best
fit to the transverse correlation function in \S~\ref{sec:discuss_size}.

\subsection{Verification of the Line-of-Sight Correlation Function} \label{sec:cf_los}

\begin{figure}
 \hbox{
  \includegraphics[scale=0.35,angle=-90,clip=true]{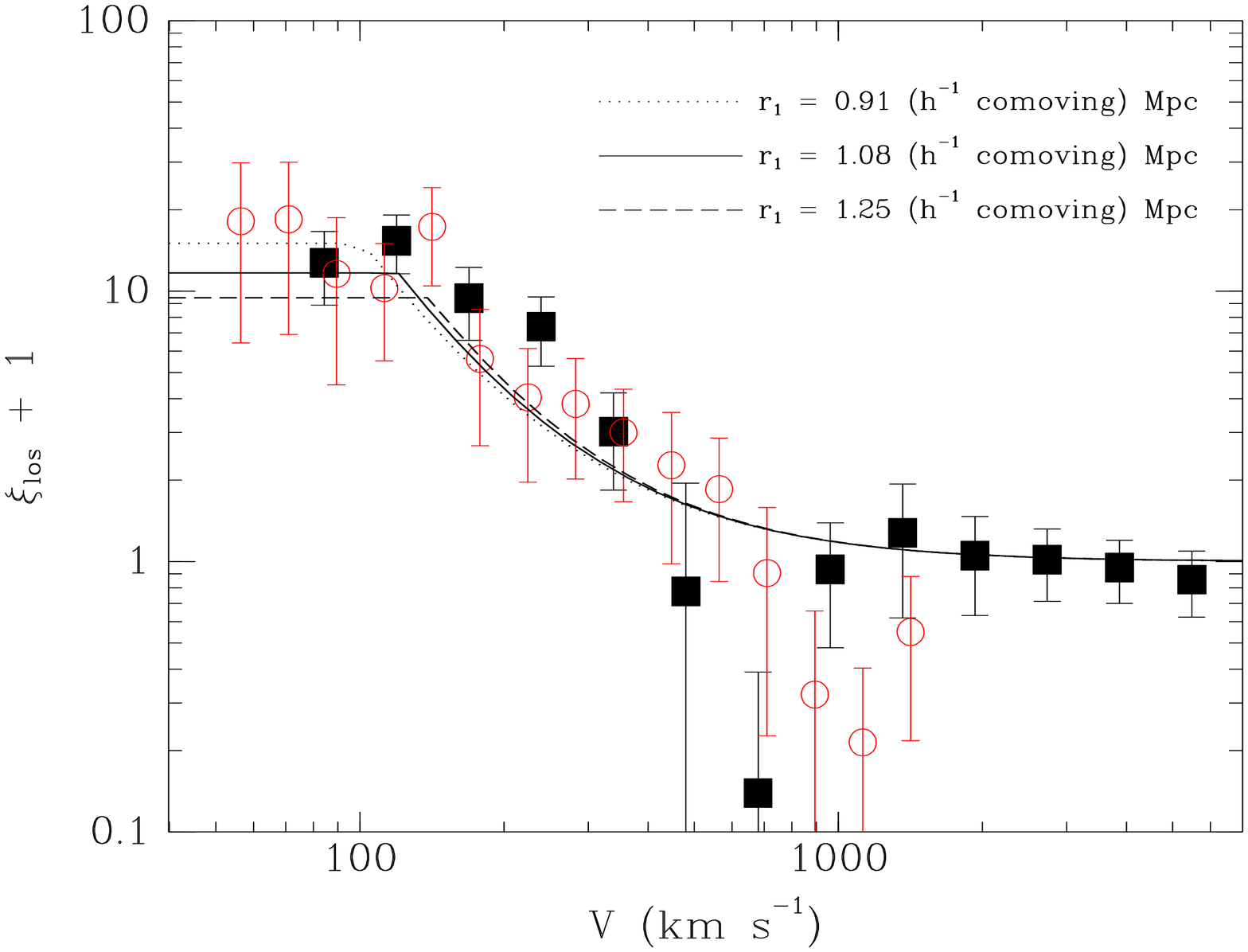}
   \hfill}
    \caption{Two-point correlation function of \cIV absorption systems
      vs. line-of-sight distance. The new ESI data (squares) agree with 
      the earlier VLT data (red circles).  The \cIV components  
      shown in Figure 10 of Scannapieco \et (2006) were grouped into
      systems using a linking length of 50\kms.       The 
      bins near 500\kms have a small number of systems; and we 
      attribute the low pair counts to systematic error introduced
      by the \cIV doublet spacing.  Fixing $r_0 = 4.0$\hinv comoving Mpc
      and $\gamma = 1.6$ in Eqn.~\ref{eqn:fit_function}, we fit 
      $r_1 = 1.08 \pm 0.17$\hinv comoving Mpc to the joint data (solid line);
      the dashed and dotted lines illustrate the 1-sigma error in $r_1$.
}
       \label{fig:xi_los} \end{figure}

For each quasar spectrum (and fake spectrum), we calculated the velocity separations 
$\Delta v$ of the \cIV systems. The separations of data-data, data-random, and random-random
pairs were binned. When we sum over all sightlines $l$, the counts yield the correlation strength
\begin{eqnarray}
\xi_{LOS}(\Delta v)  =  \frac{\sum _l {\dada} -2\sum _l{\dara} + \sum _l{\rara}
 }{\sum _l {\rara}}.
\end{eqnarray}
As discussed previously for $\xi_{AB}$, this estimator has a larger error than 
the Poisson term when $\xi > 1$. Appendix~A
describes our error estimate. These errors are shown on our correlation function in
Figure~\ref{fig:xi_los}. 
The clustering amplitude grows steadily as the velocity difference of the pairs 
decreases from 500 to 100\kms. Below separations of about 100\kms,  the 
clustering strength is constant. Our measurement confirms the break
described by Scannapieco \et (2006a). Their spectra, obtained at higher
spectral resolution with UVES, demonstrated the near constant amplitude
down to the linking length.

We fit the line-of-sight correlation with a simple model. 
Any model of the auto-correlation function for \cIV systems must break 
at small separations to fit the relatively flat correlation amplitude at 
small velocity separation. Scannapieco \et (2006a) estimated a break 
at 150\kms, corresponding to 1.34\hinv cMpc.
We convert velocity differences into
a redshift-path length, $r = \Pi = (1+z)\Delta v / H(z)$, using the
median redshift of \rr pairs in each bin. 
We assume a real-space correlation function of the form
\begin{eqnarray}
\xi(r) & = &  (r_0/r)^{\gamma} {\rm ~for~ } r \ge r_1  \nonumber \\
       &   & (r_0/r_1)^{\gamma} {\rm ~for~ } r < r_1. 
\label{eqn:fit_function} \end{eqnarray}
The redshift-space correlation function follows from Eqn. 15 of 
Hawkins \et (2003), which estimates the perturbation caused by
linear infall. Due to the slope of the correlation function,
infall lowers the amplitude of the line-of-sight correlation
function at separations much larger than $ r_1$. At smaller separations 
where the real-space correlation amplitude is constant, the volume effect
caused by infall again boosts the correlation amplitude (as it did for the
transverse correlation function).

The best fit to the joint
ESI plus UVES data yields $r_0 = 4.9 \pm 0.7$ and $r_1 = 1.2 \pm 0.3$ for
$\gamma = 1.6$. This correlation length and power-law index
are consistent with the values  $r_0 = 4.0 \pm 0.6$\hinv comoving Mpc
and $\gamma = 1.6 \pm 0.1$ fitted to the galaxy-galaxy correlation function 
at $z=2.9$ (Adelberger \et 2005a). Fixing $r_0 = 4.0$\hinv comoving Mpc 
and $\gamma = 1.6$, we fit $r_1 = 1.08 \pm 0.17$\hinv comoving Mpc
to the joint ESI plus UVES line-of-sight \cIV correlation function.  This
fit and the 1-sigma uncertainty range are plotted in \fig~\ref{fig:xi_los}.
They provide a reasonable description of the data. However,
the correlation length $r_0$ strongly influences the fitted
break at $r_1$; and we examine the implications of this parameter degeneracy
in the next section.

\section{DISCUSSION}  \label{sec:discussion}

On scales larger than the typical virialized halo, the correlation
function of galaxies is expected to represent the distribution of dark
matter halos that host them. A larger correlation length $r_0$ corresponds
to a greater probability of finding a neighbor. Populations that cluster 
more strongly have higher $r_0$ and reside in more massive halos. From
the cross-correlation of Lyman-break galaxies with \cIV systems, Adelberger
\et (2005b) showed that \cIV systems with column density $\sgreat\ 10^{13}$\col\ 
cluster as strongly as Lyman-break galaxies. It follows that these metals
and Lyman-break galaxies populate the same halos. Tens of thousands of 
Lyman-break galaxies  have been identified photometrically, 
and many thousand observed spectroscopically. It therefore seems reasonable 
to assume that the single power law that describes the clustering of 
Lyman-break galaxies on scales greater than 1\hinv comoving Mpc also 
describes \cIV systems on large scales. On scales smaller than virialized
halos, multiple galaxies orbit within a deeper potential well; and the
clustering of galaxies is harder to predict due to dissipational processes.

Our interest here is the scale of metal enriched regions $r_1$. On scales
$r < r_1$, the circulation of metals by feedback processes smoothes out their 
distribution; and the \cIV auto-correlation amplitude is roughly constant.
A difference between the fitted value of $r_1$ in redshift space and in real space 
is of particular interest. It may reflect the Doppler shifts of enriched regions. 

\begin{figure}
 \hbox{
  \includegraphics[scale=0.4,angle=0,clip=true]{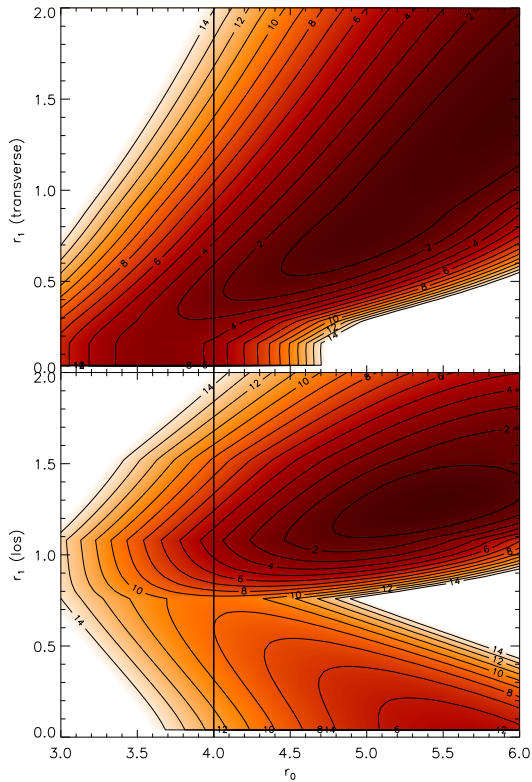}
   \hfill}
\caption{
Contours of $\Delta \chi^2$ for two-parameter fits to the \cIV correlation 
function. The correlation length $r_0$ and the size of enriched regions $r_1$ 
parameterize Eqn.~\ref{eqn:fit_function}, and the power-law index is taken
to be $\gamma = 1.6$. The fitted transverse and line-of-sight \cIV correlation 
functions are illustrated separately in the upper and lower panels, respectively. 
The contours indicate values of $\Delta \chi^2$, where
2.30, 4.61, and 6.17 correspond to the 68.3\%, 90.0\%, and 95.4\% confidence level.
{\it The line-of-sight and transverse correlation functions sample the same enriched
regions, so they must be jointly fit with the same $r_0$ and $r_1$.} 
Adopting the clustering length of Lyman-break galaxies (vertical line) implies
the presence of peculiar velocities among \cIV systems. Larger values of the 
correlation length require smaller velocities. No Doppler shifts are required
when the correlation length increases to (the rather large value of) 
6\hinv comoving Mpc and $r_1 \simeq 1.3$\hinv comoving Mpc.
(The fits to the line-of-sight clustering bifurcate at  $r_0 \sgreat 4.5$\hinv 
comoving Mpc due to errors in counting \cIV systems  separated by the doublet 
spacing, $\Delta v = 500$\kms.) 
}
\label{fig:contour} \end{figure}

For the \cIV systems, the fitted $r_1$ depends on the correlation length.  As $r_0$ 
increases, the clustering amplitude at any separation $r > r_0$ increases. To
remain consistent with the measured correlation strength at small separations, 
the $r_1$ transition must shift to larger scales, where the correlation
amplitude is lower. Figure~\ref{fig:contour} shows how the fitted $r_1$ grows 
with increasing correlation length. Based on the shape of the contours, the 
line-of-sight measurements clearly constrain the scale $r_1$ more tightly 
than they do the correlation length $r_0$. The difference in $r_1$ values
and their implied Doppler shifts decrease as the correlation length increases.

The underlying real-space correlation length $r_0$  must describe both the 
transverse and line-of-sight clustering measurements. Clustering along the 
line-of-sight provides the better constraint  because our sightlines to the 
binary quasars were simply not far enough apart to constrain $r_0$ well. 
Both measured \cIV correlation functions, nonetheless, are marginally 
consistent with the Lyman-break galaxy correlation length, vertical line 
at  $r_0 = 4.0 \pm 0.6$\hinv comoving Mpc in Figure~\ref{fig:contour}.
The best-fit correlation length $r_0 = 4.9 \pm 0.7$\hinv  is slightly larger.
In the absence of any peculiar velocities, the correlation length must be 
increased to $ 6$\hinv comoving Mpc before the same $r_1$ fits the clustering 
measured  in {\it both} redshift-space {\it and} real space.  Our \cIV 
observations provide only weak constraints on very large-scale clustering,
so they alone do not rule out  $r_0 \simeq 6$\hinv comoving Mpc.  
However, such halos are about three times more massive than those surrounding 
Lyman-break galaxies and have a number density about 4 times lower.\footnote{
      At $z = 2.0$, Lyman-break galaxies have $M_h = 10^{11.2 - 11.8}$\msun\ 
      (Adelberger \et 2005a)} 
In light of the strong cross-correlation amplitude between \cIV systems and 
galaxies (Adelberger \et 2005b) and the presence of outflows from most 
Lyman-break galaxies (Steidel \et 2010), we do not discuss this possibility further.

For $r_0 = 4.0$ or 4.9\hinv
comoving Mpc, the implied peculiar velocities set an upper limit on the 
average speed of galactic outflows. Together, the size and speed measurements 
place interesting new constraints on the era of metal dispersal.

\subsection{The Size of Metal-Enriched Regions} \label{sec:discuss_size}

\subsubsection{Amplitude of the Transverse Correlation Function}

\begin{figure}
 \hbox{
  \includegraphics[scale=0.5,angle=-90,clip=true]{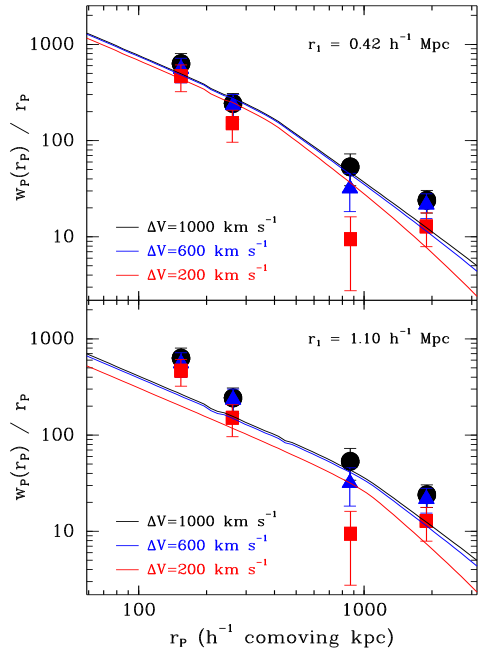}
   \hfill}
    \caption{
      Model of the correlation function defined by Equation~\ref{eqn:fit_function}
      in real-space and then corrected for linear infall using Equation~15 of 
      Hawkins \et (2003).    We fix the correlation length at
      $r_0 = 4.0$\hinv comoving Mpc and power-law index at $\gamma = 1.6$,
      (Adelberger \et 2005a).
      We illustrate how the scale $r_1$ affects the transverse 
      correlation amplitude. 
      Data are as shown in Fig.~\ref{fig:w_dv_data}.
      (a, Top) A fit to these data (specifically $\Delta v = 100$\kms )
      suggests $r_1 = 0.42 \pm 0.15$\hinv comoving Mpc. We
      show this model and interpret $r_1$ as 
      the typical radius of an enriched region.
      (b, Bottom) 
      We show a model with $r_1 = 1.08$\hinv comoving Mpc, as favored by
      our fit to the line-of-sight correlation function.
      The larger size for the enriched regions provide too little power on 
      small scales. }
     \label{fig:w_model_rf} \end{figure}

As our best estimate for the size of metal-enriched regions, we adopt 
$r_0 = 4.0$\hinv comoving Mpc and  $\gamma = 1.6$ (Adelberger \et 2005a). 
We fit the model given by Equation~\ref{eqn:fit_function} to the transverse 
correlation function and find the power-law flattens on a scale 
$r_1 = 0.42 \pm 0.15$\hinv comoving Mpc. This fit is shown in the top panel of 
Figure~\ref{fig:w_model_rf}, where it provides a good description of
our measurements. For $r_0 = 4.0$\hinv comoving Mpc, the fit to the 
line-of-sight correlation function shown in \fig~\ref{fig:xi_los} 
requires larger $r_1 = 1.08 \pm 0.17$\hinv comoving Mpc. A model illustrating
this larger transition scale is shown in the bottom panel of 
Figure~\ref{fig:w_model_rf}, where it clearly underpredicts the correlation 
between sightlines seen on small scales. We find \cIV systems at similar 
redshifts in the two sightlines more often than we would in a  model 
with metal bubbles as large as those suggested by Scannapieco \et (2006a).
Consistency requires a model with redshift-space distortions.

We also illustrate the discrepancy in redshift-space when the \cIV correlation 
length takes our best-fit value of $r_0 = 4.9$\hinv comoving Mpc for 
$\gamma = 1.6$. We fitted  $r_1 = 1.21 \pm 0.25$\hinv comoving Mpc to the line-of-sight 
correlation function. To the transverse measurement, we fitted $r_1 = 0.74 \pm 0.21$\hinv 
comoving Mpc. The scale of the line-of-sight break remains 
significantly larger than the fitted transverse break. Evidently, 
the metal-enriched regions present peculiar velocities with
respect to the cosmic expansion. We fit the magnitude of these
velocity kicks in \S~\ref{sec:discuss_v}.

\subsubsection{Coherence Length from Coincident Absorption in Spectra of Binary Quasars} \label{sec:coherence}

The spectra of binary or lensed quasars have been used previously to constrain
the transverse size of absorbing structures. The transverse size of
a coherent region was inferred to be at least as large as the maximum separation 
of common spectral features  (McGill 1990; Lopez \et 2000; Ellison \et 2004). As
implemented previously, however, the resulting coherence length has little to
do with the size of enriched regions. To illustrate this important point, we
apply the standard coherence-length analysis to our sample of \cIV systems,
explain why it fails, and suggest a revised technique. This revised coherence-length 
estimator can be compared to the size scale
$r_1 = 0.42$ or 0.74\hinv comoving Mpc fitted to the transverse correlation function.

\begin{figure}
    \includegraphics[scale=0.35,angle=-90,clip=true]{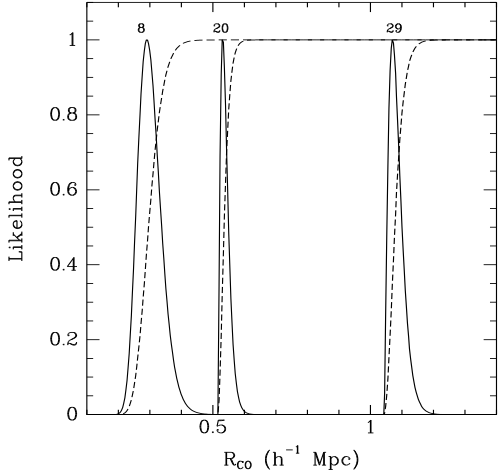} 
  \caption{Likelihood of observed coincident and anti-coincident \cIV systems 
    (normalized by the maximum likelihood)  vs. the coherence length of a spherical 
    distribution of clouds.  
    The dashed line shows the cumlative distribution, which we use 
    to define the 95\% confidence limit on the statistical error. We    
    defined {\it coincidences} as systems in different sightlines within 600\kms of each other. 
    The most likely coherence radius $R = 1.07^{+0.07}_{-0.02} $\hinv Mpc
    for all 29 pairs. However, the best radius drops to $R \simeq 0.5$\hinv Mpc when the
    9 widest binaries are dropped from the sample; and $R \simeq 0.3$\hinv Mpc when
    the analysis is applied to only the closest 8 binaries in the sample. As sample
    size increases (and wider binaries are included), the maximum-likelihood analysis
    does not converge to a coherence length. See text for explanation.
   }
  \label{fig:like_data} \end{figure}

To estimate the coherence scale of \cIV absorption, we first follow the formalism of 
McGill (1990) with the modification suggested by Dinshaw \et (1997). Their argument
is entirely geometric in nature. They describe absorption-line systems with a 
non-evolving population of clouds. For illustration, we adopt a spherical distribution of clouds, but the 
method easily generalizes to cylindrical filaments or inclined disks. The radius $R$
describes the coherence scale of the cloud distribution. 
For pairs of sightlines separated by a distance $l$,\footnote{ 
    Although previous authors used the physical separation of the sightlines,
    the co-moving separation must be used when the absorber redshifts span a
    significant range in the growth factor, $1+z$.}
the probability that the second sightline 
intersects the sphere, given  that the first sightline ``hit'' the sphere is
\begin{eqnarray}  \label{eqn:prob}
P_1  = 2/\pi \{ \cos^{-1} X(z)  - X(z) \sqrt{1 - X(z)^2} \},
\end{eqnarray}
where $X(z) = l(z) / (2R)$ and $ 0 \le X(z) \le 1$.
If {\it either} sightline shows absorption, the probability of coincident
absorption, defined as \cIV absorption in both sightlines at similar redshift,
is 
\begin{eqnarray}
P_2  = P_1 / (2 - P_1).
\end{eqnarray}
The probability of absorption in just one sightline, an anti-coincidence, 
is $1 - P_2$. For a particular coherence length $R$, the likelihood of obtaining 
the observed number of coincident systems {\it and} the observed number of anti-coincidences
is
\begin{eqnarray} \label{eqn:like}
{\cal L} (R) = \Pi_{i=1}^{N_i} P_2[X(z_i)]  \Pi_{j=1}^{N_j} (1-P_2[X(z_j)]),
\end{eqnarray}
where the two products are over all coincidences $N_i$ and anti-coincidences $N_j$,
respectively. Figure~\ref{fig:like_data} shows the likelihood for many different 
values of the coherence length. 


Applied to the entire sample of 29 binary quasars, the maximum likelihood is found 
at the very large scale of $R = 1.07^{+0.07}_{-0.02}$\hinv comoving Mpc. Detection
of just a single coincidence between sightlines separated by $l =2.1$\hinv
comoving Mpc sets a sharp lower bound. In general, the analysis proposed
by McGill (1990) and Dinshaw \et (1997) requires the sphere diameter to be at least
as large as the separation of the widest pair of sightlines with a coincidence.
As we argued at the beginning of \S~\ref{sec:discussion}, however, the sources
of metals are highly clustered.  When two sightlines are separated by $\sim 1$\hinv 
comoving Mpc, the correlation between the halos of their sources causes
some absorption at a common redshift.
The clustering of galaxies, and therefore metal-enriched regions, has
not been accounted for in Eqn.~\ref{eqn:prob}. This oversight leads to
a coherence scale that grows with the separation of the widest binary quasar
in the sample.  Figure~\ref{fig:like_data} demonstrates this lack of
convergence by comparing the likelihood functions for three subsets of
data.

\begin{figure}[h]
{ \hbox{
         \includegraphics[scale=0.35,angle=-90,clip=true]{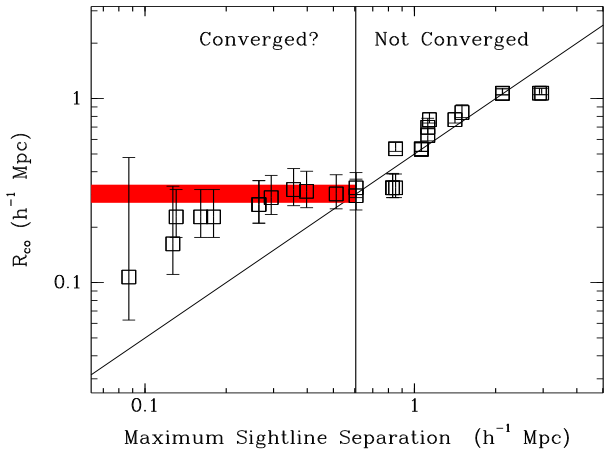}
          \hfill}}
           \caption{
             Coherence length (in comoving units) for different subsets of paired
             sightlines. Using our large set of pair observations 
             at varying separations,  we iteratively threw out the largest pair and recomputed
             the coherence length using the maximum likelihood analysis. 
             The $x-axis$ inidicates the widest (comoving) separation
             included in the subsample. The diagonal line indicates a sphere radius
             of one-half the maximum sightline separation, $R = 0.5 l$. 
             The estimated coherence length does not converge as the sample
             size is increased. When $R$ is significantly greater than $0.5l$,
             however, this method may provide a robust estimate of the
             coherence length (shaded, red) of individual enriched regions.
}
       \label{fig:converge} \end{figure}

In principle, it might be possible to modify the analysis to obtain a robust
coherence length for enriched regions. This provides no obvious advantage over 
the \cIV correlation function but offers a sanity check on the inferences drawn 
from the correlation function. We suggest that the maximum-likelihood estimate
of the coherence length may converge provided the sightline separations remain
smaller than the typical size of individual enriched regions.   When the likelihood
analysis yields $R$ comparable to one-half the maximum sightline separation, 
the method has not converged. To estimate
this coherence scale, we calculated the function given by Eqn.~\ref{eqn:like}
for a series of 1, 2, 3, ..., and 29 binary quasars. Figure~\ref{fig:converge} compares
our result to the line $2R = l$. When the sightline separation is smaller 
than the diameter of the typical enriched region, $l < 2R$, we expect many coincidences, 
few anti-coincidences, and a most-likely coherence diameter $2R$ significantly larger 
than the maximum sightline separation.

In Figure~\ref{fig:converge}, the coherence length estimate for
all subsamples with maximum sightline separation less than 0.6\hinv comoving
Mpc allows $R \approx 0.3$\hinv comoving Mpc. In contrast, as the maximum sightline
separation grows larger than 0.6\hinv comoving Mpc, the estimated coherence length
follows the locus $R = 0.5 l$. The coherence length appears to converge
to a value $R \approx 0.3$\hinv comoving Mpc in Figure~\ref{fig:converge}.  
This scale is consistent with
the $0.42 \pm 0.15$ \hinv comoving Mpc fitted to the transverse correlation function
with $r_0 = 4.0$\hinv comoving Mpc. It is marginally inconsistent with the larger $r_1$ 
fitted when a larger correlation length is adopted. 
Given the many implicit assumptions in the interpretation of both
$R$ and $r_1$, and the acknowledgement that enriched regions could have a range
of sizes, we believe the revised coherence length analysis lends strong support
to our proposed interpretation of the \cIV correlation function. 
It may rule out $r_1$ as large as 0.7\hinv comoving Mpc, but we prefer to
see the technique calibrated against cosmological models before making this claim.

\subsubsection{Previous Observations of \cIV Absorption near Lyman-Break Galaxies}

Sightlines passing by known Lyman-break galaxies directly probe the
properties of circum-galactic gas. Spectra of background light beacons present
intervening absorption near the redshift of Lyman-break galaxies. 
Because background galaxies are numerous, spectroscopy of galaxies can
probe circum-galactic gas at low impact parameter. Background quasars are 
less common, but spectra of the brightest quasars provide higher sensitivity 
to absorption lines. The impact parameters of these detections can be compared
to our measured scale for the enriched regions, about 150 physical kpc.

Composite galaxy spectra were recently created over a range of
impact parameters, $b$, from Lyman-break galaxies (Steidel \et 2010). 
Intervening \cIV absorption was detected at  $b=60$~physical~kpc
but not at $b = 100$~kpc. Our analysis suggests the radius of the 
typical metal-enriched region is significantly larger. Is this
a sensitivity difference or a discrepancy?

Our sample contains \cIV systems as strong as the $W_{r,1549} = 700\mA $
absorption trough in the composite spectrum at $b=60$~kpc. Our spectra
of median quality are more sensitive than their
detection threshold of $W_{r,1549} = 120$\mA\ at  $b = 100$~kpc. 
Figure~\ref{fig:few} illustrates our equivalent width distribution.
The typical detection limit is $W_{r,1551} = 
0.5 W_{r,1549} \simeq 20$\mA; and a few spectra reach
$W_{r,1551} = 0.5 W_{r,1549} \simeq 8$\mA. Our sensitivity to weaker systems 
may explain why we find a larger size for the enriched regions. 

These weaker  \cIV systems do appear to be associated with the
environment of Lyman-break galaxies. The clustering of weaker systems
has been measured using sensitive spectra of 23 quasars in Lyman-break galaxy 
survey fields. The cross-correlation length (between LBGs and \cIV systems) 
increases slowly with the \cIV column  density (Adelberger \et 2005b, Figure~11).
At the completeness limit of our survey, $W_{r,1551} \simeq 20\mA$, 
 \cIV systems cluster as strongly with Lyman-break galaxies as
Lyman-break galaxies do with each other.
Hence, we can be reasonably confident that the large metal enriched
regions describe the environment of Lyman-break galaxies.
We have not shown, nor do we claim, that these weaker \cIV systems
mark outflows from Lyman-break galaxies.

\subsection{The Peculiar Velocities of Metal-Enriched Regions} \label{sec:discuss_v}

It is clearly of interest to understand why the line-of-sight correlation
between \cIV systems flattens on a larger scale than does their transverse 
correlation function. The line-of-sight correlation function is derived
from the redshift difference of pairs of \cIV systems. Their separation
in redshift mixes information about their physical separation and their relative
Doppler shifts. Non-zero peculiar velocities smear out the correlation
in redshift space. Adding random velocity kicks to enriched regions
therefore flattens the line-of-sight correlation function on a scale
somewhat larger than $r_1$. We can fit the magnitude of the typical 
velocity kick.

\begin{figure}
  \includegraphics[scale=0.35,angle=-90,clip=true]{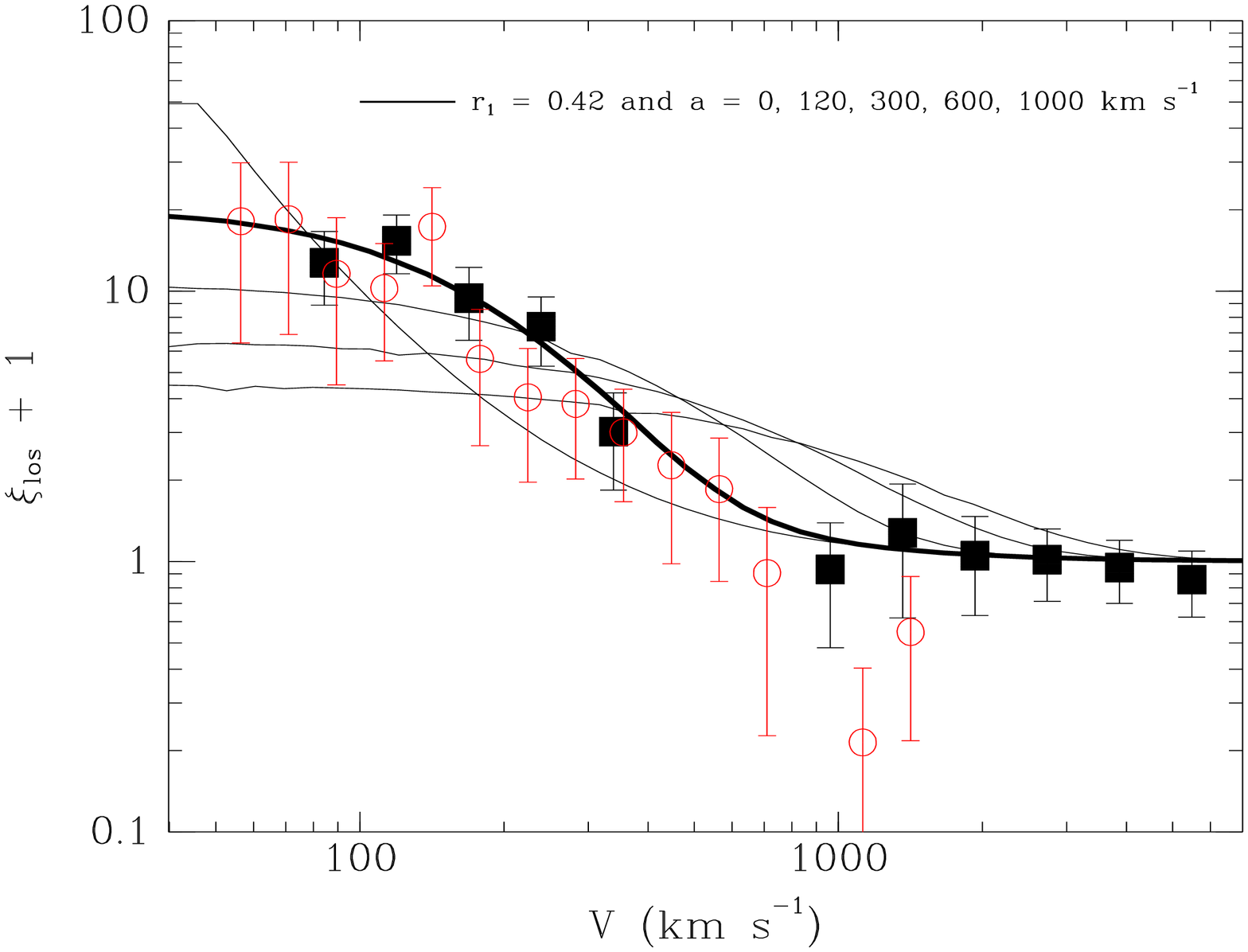}
  \includegraphics[scale=0.35,angle=-90,clip=true]{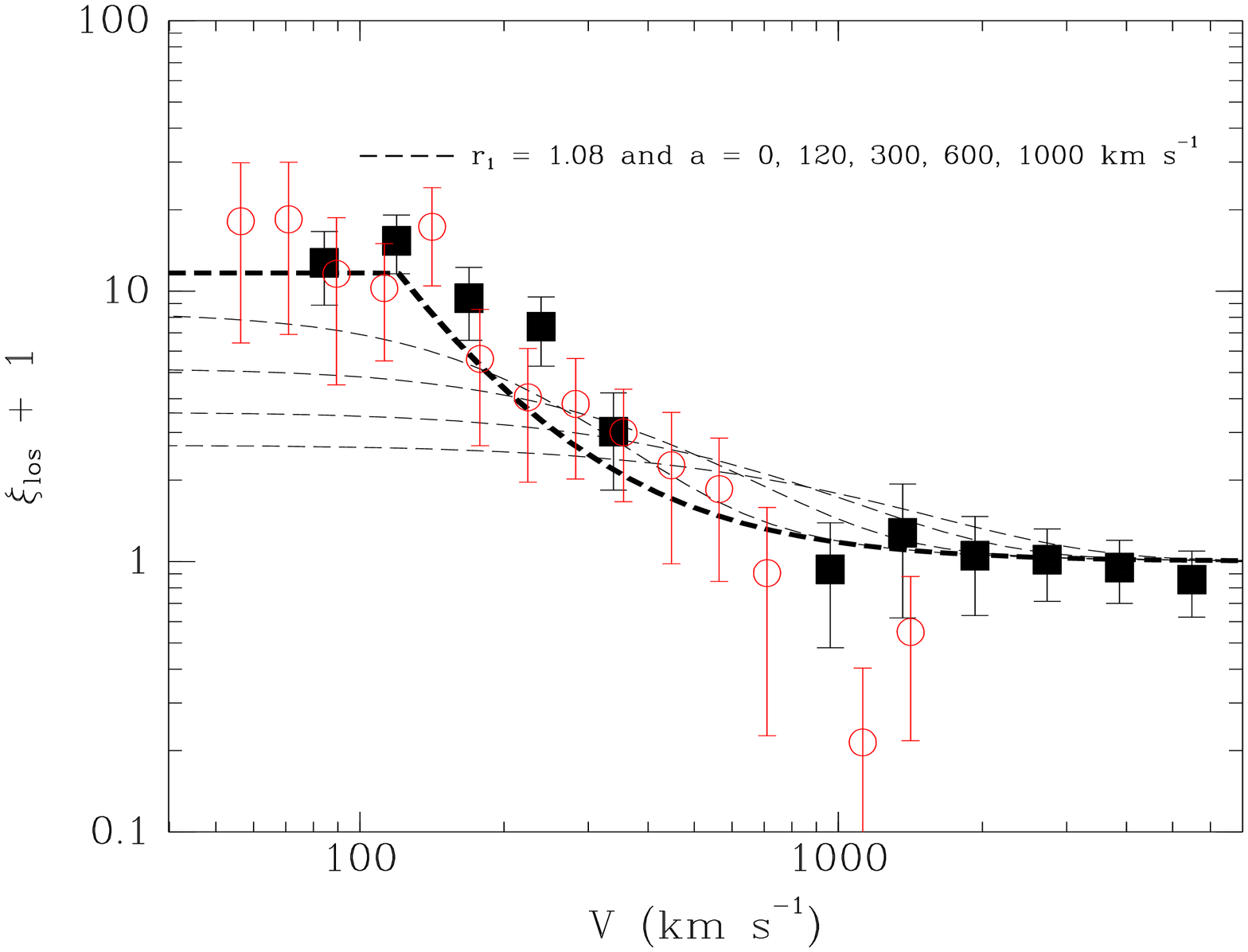}
    \caption{Models for the line-of-sight correlation function.
      The parameter {\it a} in Eqn.~\ref{eqn:peculiar} describes
      the typical random velocity of an enriched region. The
      real-space correlation function and infall models are
      the same as those shown in Fig.~\ref{fig:w_model_rf}
      (with $r_0=4.0\hinv$ comoving Mpc and $\gamma = 1.6$).
        Data are as in \fig~\ref{fig:xi_los} with the bins affected by the
        doublet spacing omitted for clarity. (Top) We fix $r_1 =
        0.42$\hinv comoving Mpc, the value favored by our fit to the transverse
        correlation function. Thick line shows the best
        fit peculiar velocity $a = 120$\kms. Thin lines illustrate
        other peculiar velocities. Lower
        velocities give too much correlation on small scales; and higher
        velocities require more  pairs at velocity separations $\Delta v \approx 300$
        to 1000\kms than are observed.
        (Bottom) Here we fix $r_1 = 1.08$\hinv comoving Mpc, the
        fitted bubble size for no peculiar velocity (bold line).
        We show that adding even a small peculiar velocity of $a=120$\kms
        results in too few pairs of \cIV systems at $\Delta v < a$
        and too many pairs at $\Delta v > a$.
        }
        \label{fig:xi_los_gal}  \end{figure}

\begin{figure}
 \hbox{
  \includegraphics[scale=0.35,angle=-90,clip=true]{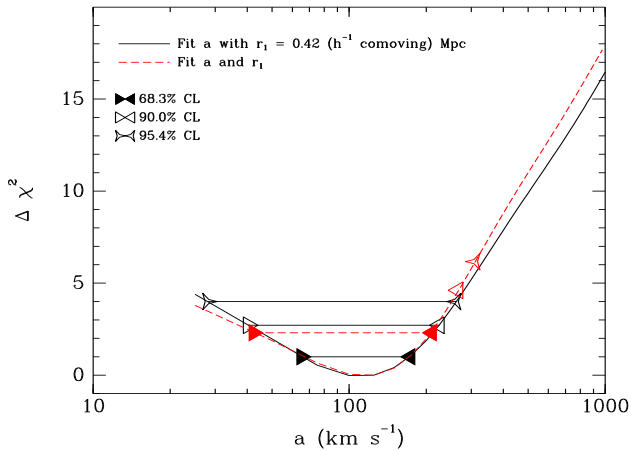}
   \hfill}
    \caption{Values of peculiar velocity $a$ fitted to the
      line-of-sight \cIV correlation function.
      With $r_1$ fixed at 0.42\hinv comoving Mpc (solid line),
      the deviation in $\chi^2$ from the best fit allows
      peculiar velocities up to 170, 220, and 260\kms at
      the 1, 2, and $3\sigma$ confidence level. A more
      general 2-parameter fit to $a$ and $r_1$ simultaneously 
      (dashed line) allows a maximum peculiar velocity of 310\kms
      at the $3\sigma$ level. (The correlation function
      of Adelberger \et (2005a) is assumed for $r_0$ and $\gamma$.)
 }
     \label{fig:a} \end{figure}

As suggested by Hawkins \et (2003), we represent the random motions by an exponential 
of the form 
\begin{eqnarray}
f(v) = \frac{1}{a\sqrt{2}} \exp\left( - \frac{\sqrt{2} |v|}{a}  \right).
\label{eqn:peculiar} \end{eqnarray}
The impact of the Doppler velocity $a$ on the line-of-sight correlation function
is shown in \fig~\ref{fig:xi_los_gal}, where these kicks increase from zero to 
$a = 1000$\kms.  The correlation amplitude decreases at velocity separations $\Delta v < a$
and grows on the  scale $\Delta v \simeq a$. 
The magnitude of the velocity kick required to fit the correlation function
depends on $r_1$, $r_0$, and $\gamma$. The largest value of $a$ results if
we adopt the smallest value of $r_1$ (and therefore the smallest correlation 
length). For example, taking $r_1= 0.42$\hinv comoving Mpc as shown 
in the top panel of \fig~\ref{fig:xi_los_gal}, we fitted $a \simeq 120$\kms -- the model
shown by the bold line. With no velocity kick, we fitted a larger
$r_1 = 1.08$\hinv comoving Mpc to these data in \fig~\ref{fig:xi_los}.
Models with these larger enriched regions do not allow any velocity kicks. As
illustrated in the bottom panel of \fig~\ref{fig:xi_los_gal}, large enriched
regions with peculiar velocities yield too few pairs of \cIV systems at separations 
of a few hundred \kms.  To summarize, we quantify the upper limit on $a$  in \fig~\ref{fig:a}. 
Varying the Doppler velocity and computing the fit statistic, we find 
a $3\sigma$ upper limit  $a \le 300$\kms. {\it We conclude that
the average Doppler shift of metal-enriched regions along the line-of-sight 
is likely about 120\kms and appears very unlikely to exceed 300\kms.}

The average peculiar velocity of metal-enriched regions in three-dimensions is $v_{pec} = 
\sqrt{3} a \simeq 200$\kms ($3\sigma$ upper limit of 500\kms) after correction for 
projection along the line-of-sight. 
These peculiar velocities are similar to those of galaxies. Their primary source 
may be gravitational instability associated with the underlying mass structure. 
The outflow speed measured in galaxy spectra is $\simeq 200$\kms
(Martin 2005; Weiner \et 2009; Steidel \et 2010).
This absorbing gas lies close the host galaxy (Martin \& Bouch\'{e} 2009).
The average outflow speed, where the average is a spatial average from the 
starburst region out to the bubble radius where the wind stalls or even turns around, 
has not been directly measured previously. {\it Our measurement shows that this average
outflow speed is very unlikely to be larger than 200\kms.} The average
velocity may be  considerably less because gravitationally-induced peculiar 
velocities must contribute to random velocity kicks.

Adelberger \et (2005b) reached a very different conclusion about average
outflow speeds. Their Figure~12 shows the  distribution of redshift 
differences between \cIV systems and Lyman-break galaxies. If the
galaxy and gas are co-spatial, the typical offset in redshift 
corresponds to a Doppler shift of $\simeq 500$\kms. Because such
velocities are larger than virial motions within Lyman-break galaxies,
this observation was interpreted as evidence for galactic outflows
at 500\kms or more.

Outflow speeds of 500\kms are inconsistent with the measured line-of-sight 
correlation function for \cIV systems. We suggest that the redshift 
differences between \cIV systems and Lyman-break galaxies can be 
understood without appealing to fast outflows. The physical separation
between the absorbing gas and the galaxy may explain their redshift
differences. For example, cosmic expansion at $z \simeq 3$
produces a velocity difference of 450\kms over the 4.0\hinv comoving Mpc
(the correlation length). As the column density of \cIV systems increases, 
their cross-correlation 
length (with galaxies) increases.  For larger cross-correlation length,
which is equivalent to a stronger correlation amplitude, the 
distance from a galaxy to the nearest intervening systems is smaller.
The average column density of the absorption will be higher the 
smaller the separation or impact parameter.
The strongest \cIV sytems are always found at small impact parameters,
so random spatial offsets between galaxies and \cIV systems can 
plausibly explain the increase in  redshift differences observed with 
growing impact parameter.

\subsection{The Dispersal of Metals} \label{sec:discuss_origin}

Understanding how \cIV systems cluster reshapes our view
of the IGM enrichment history. In particular, the large sizes and low peculiar
velocities of enriched regions constrain the era of metal dispersal.
This process is of course on-going as there is strong observational evidence for 
galactic winds across much of cosmic time. Not all galaxy populations, however, can 
easily distribute metals as widely as observed.

\subsubsection{Lyman-break Galaxies at $z \simeq 3$}

How far can a galactic wind from an individual galaxy can transport metals? 
In the absence of any forces to stall the expansion, the maximum distance traveled
is limited by the age of the starburst galaxy and average velocity. Adopting
the upper limit  $\langle v \rangle \simeq 200$\kms compatible with line-of-sight 
clustering observations, galactic winds travel
\begin{eqnarray}
R \simeq 61 {\rm ~kpc~} 
\left( \frac{ v }{ 200 {\rm ~km s}^{-1}}  \right)
\left( \frac{\tau}{300 {\rm ~Myr}}  \right) 
\end{eqnarray}
in a typical LBG lifetime.\footnote{For purposes of illustration, we use the 
             median timescale for star formation, 321~Myr, from Table 4 of Shapley \et (2001). 
             With only optical photometry, the absolute ages
             are quite sensitive to the model for the star formation history.  For example,
             Sawicki \& Yee (1998) fit much younger ages $\sim 30$~Myr. As shown by
             Shapley \et (2001), measurement of
             the Balmer break distinguishes these interpretations and favors
             typical ages of a few hundred Myr.}
This distance is at most 40\% of our estimated radius for an
enriched region. While typical Lyman-break galaxies clearly eject metals in winds, 
other galaxies may spread metals over larger regions of intergalactic space.

The properties of Lyman-break galaxies do span a signficiant range;
and the winds from those with the largest lifetimes can enrich the
large regions we find. To disperse metals out to 200~kpc, the required age
is $\sim 1 $~Gyr  for an average speed $\simeq 200$\kms. Ages $\sim 1 $~Gyr 
are fit to about 20\% of Lyman-break galaxies (Shapley \et 2001).
Interestingly, the Lyman-break galaxies with older ages tend to be less
luminous (Shapley \et 2001). It follows that the lowest luminosity Lyman-break 
galaxies (in current samples) could be the source of the most dispersed metals.

Given the absorption cross-section of bubbles with $r_1 \simeq 0.42$\hinv comoving Mpc,
the Lyman-break galaxies are not numerous enough to be the source of most
of the \cIV systems. The number density of Lyman-break galaxies is $n_{g} \approx 3.7 
\times 10^{-3} {\rm ~h}^3$\mpc-3 (Steidel \et 2010). If we could place these bubbles
around every Lyman-break galaxy, the resulting redshift-path density of \cIV 
systems would be about
\begin{eqnarray} 
\frac{dN}{dz} \approx 1.4 \left ( \frac{n_{gal}}{3.7 \times 10^{-3} h^3 {\rm ~Mpc}^{-3}} \right ) 
\times \nonumber \\
\left ( \frac{r_1}{0.42 \hinv {\rm ~Mpc}} \right )^2
\left ( \frac{450 h {\rm ~km~s}^{-1} {\rm Mpc}^{-1}}{H(z)} \right ).
\label{eqn:dndz} \end{eqnarray}
For comparison, integration of our distribution function shown in
\fig~\ref{fig:few} to 
$W_{r,1551} \simeq 20 \mA$ yields about 14, \cIV systems per unit redshift. 
Each bubble of metals might yield more than one \cIV system, but \cIV
systems outnumber Lyman-break galaxies by an order of magnitude.

\subsubsection{Post-Reionization Dispersal of Metals}


Large bubbles of metals surrounding Lyman-break galaxies could be relics of 
an era when galaxies were smaller, more numerous, and closer together
(Porciani \& Madau 2005; Scannapieco 2005). How do the new results
on the clustering, size, and peculiar velocities of enriched regions
fit into this picture?

In order for their remnant winds to cluster like \cIV systems, higher-redshift 
sources must have lower mass. Consider the post-reionization galaxy population
for example. Redshift 6 galaxies in halos about 30 times
less massive than those of Lyman-break galaxies have the same bias as 
the Lyman-break galaxies at $z \simeq 3$. At redshift 4.3 galaxies 5
times less massive than Lyman-break galaxies would eject metals that clustered 
like Lyman-break galaxies by $z \simeq 3$.

Low-mass galaxies naturally produce large enriched regions. In a shallow
gravitational potential, gas retainment is more difficult.   Because 
enrichment is earlier, the bubbles around low-mass galaxies have more
time to grow. At an average speed of $\simeq 200$\kms, the bubble
could travel 150 physical kpc between redshift 4.3 and 3. If growth 
stalls relative to the surrounding medium, the bubbles continue to grow in 
physical size due to cosmic expansion. (For a given physical size, the comoving 
bubble scale is larger the earlier the enrichment.) Most importantly, however,
the number density of halos (and presumably galaxies) grows rapidly with decreasing 
mass. Those low-mass galaxies between $z=4.3$ and $6$ that cluster like Lyman-break
galaxies could easily be 5 to 30 times more numerous.

In the post-reionization universe, the potentially high number of galactic
winds and large cross-section for absorption yield a large density of
metal-line systems. From Eqn.~\ref{eqn:dndz}, the number per unit
redshift is about 5 when starburst galaxies at redshift 4.3 are the
source of metals observed at redshift 3. The redshift-path density
grows rapidly for earlier dispersal. The spatial distribution of 
the entire population of \cIV systems with $N(\cIV) \sgreat\ 10^{13}$\col\ 
can easily be attributed to their relic galactic winds. {\it Significant
metal enrichment will of course continue until much later, but 
the post-reionization  galaxies appear to dictate the size scale of
the circumgalactic medium.}

The bubbles from numerous low mass galaxies may overlap. Bubble coalescence
would generate rapid growth in the size of the enriched regions. Such growth 
is demonstrated by simple analytic models. In Figure~2 of Scannapieco \et (2005), 
the comoving size of enriched regions grows slowly when rare galaxies provide 
the metals.  Indeed, the bubbles blown by different Lyman-break galaxies never overlap. 
The probability of finding a second galaxy within the bubble volume $V$ is
$P = n_{gal} V (1 + \xi(r_1))$. Substituting typical values, we estimate
\begin{eqnarray}
P = 0.04 \left ( \frac{n_{gal}}{3.7 \times 10^{-3} h^3 {\rm ~Mpc}^{-3}}  \right ) \times \nonumber \\
\left( \frac{r_1}{0.42 \hinv {\rm ~Mpc}} \right )^3
\left ( \frac{ 1 + \xi(r_1)}{1 + 37} \right ), 
\label{eqn:overlap} \end{eqnarray}
where we adopt a correlation length of 4.0\hinv comoving Mpc and
$\gamma = 1.6$ to estimate 
the probability of finding another Lyman-break galaxy within 
$r_1$\hinv of the first.
Since the bubbles blown by a Lyman-break galaxy with median age are not
this large, this parameterization gives an upper limit on the odds of 
overlap. The $z \simeq 4.3$ starbursts are the youngest population (750~Myr
at $z \simeq 3$)  that have time to blow the large bubbles we found.
Their 5 times higher source density yields $P \approx 0.2$ suggesting
that many bubbles might remain isolated although approaching overlap.
In contrast, relic winds from redshift 6 sources would overlap 
create enriched regions much larger than the bubbles blown by individual galaxies.

In summary, the measured size $r_1$ of the enriched regions is difficult
to explain by metal dispersal within just a few hundred Myr of $z \simeq 3$.
This observation does not distinguish slightly earlier enrichment at
$z \simeq 4.3$ from enrichment at the end of cosmic reionization
at $z \approx 6$. Enrichment over this era, however, might be distinguished
from the Pop III scenario by measuring rapid growth in the size of
enriched regions. Pop III stars in 1000~K halos would produce rapid growth in 
the comoving size of the enriched regions prior to redshift 6. Post-reionization
wind bubbles, on the other hand, may coalesce between redshift 6 and 3,
generating rapid growth in the size of enriched regions. Our  measured
size, although large,  does not require coalescence. Starbursts at redshifts
4-5  could create such bubbles, which would not have merged for the most 
part by $z \simeq 3$.
Some of the parent galaxies would still be observable at redshift 3 and
could be identified by their older stellar population and low mass.

\subsection{Evolution of the Size of Enriched Regions} \label{sec:evolve}

The evolution of the \cIV auto-correlation function may provide a useful
constraint on the metal enrichment history. The size of enriched 
regions can only grow in comoving units if bubbles
merge or additional enrichment takes place. 
As bubbles coalesce due to 
galactic infall, the size of the enriched region grows as gravity assembles 
galaxies into large-scale structures. The correlation amplitude for galaxy 
halos grows as $(1+z)^{-2}$. Evolution in the \cIV correlation function
inconsistent with $(1+z)^{-2}$ would indicate the addition or removal of metals. 

Scannapieco \et (2006a) found that the amplitude of the line-of-sight correlation 
functions for \mgII and \feII, and possibly \cIV, increases with cosmic time. 
We used our median redshift to define lower redshift \cIV systems ($1.7 <
 z_{ab} \le\ 3 $) and higher redshift systems ($ 3 \le z_{ab} < 4.3$). 
For each subsample, we counted \dada, \dara, and \rara pairs in a single
velocity bin at $50 <  \Delta v < 500$\kms. The median pair separation
is about $\Delta v \approx 150$\kms\ at both redshifts.
The correlation amplitude of the higher redshift subsample is $\xi +1 \approx 5.5$.
The amplitude grows to $\xi + 1 \approx 11$ for lower redshift subsample.
Hence, we confirm the amplitude of $\xi_{LOS}(CIV)$, increases with cosmic time. 

\begin{figure}[h]
 \hbox{
  \includegraphics[scale=0.35,angle=-90,clip=true]{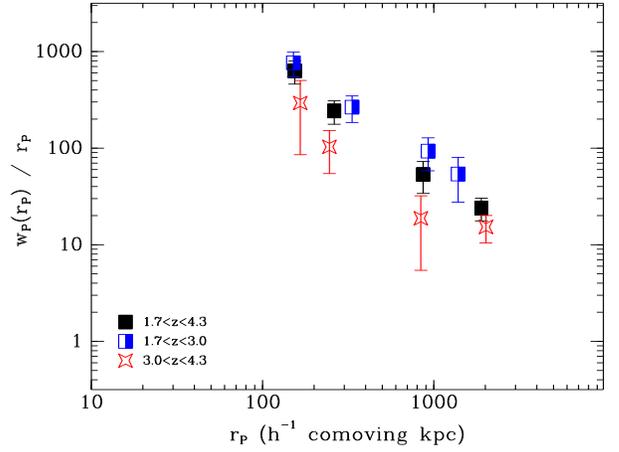}
   \hfill}
    \caption{Transverse correlation function, as in \fig~\ref{fig:w_dv_data}
        with $\Delta v \le\ 1000$\kms, but for high and low redshift \cIV systems. 
        The correlation amplitude for the lower redshift subsample at $z = 2.46$ 
        (blue, half-filled symbols) exceeds that measured for higher redshift 
        subsample at $z=3.33$ (red, starred symbols).
    }
       \label{fig:w_dz_data} \end{figure}

When we compare two sightlines, we count pairs with separations up to 
$\Delta v_{pair} = 1000$\kms. The identification of systems in different spectra 
is not biased by the \cIV doublet spacing, and the larger velocity difference
reduces the noise.
In the higher redshift subsample, the median redshift of an \rarb pair is 3.33.
The median pair redshift is 2.46 in the lower redshift subsample. We calculate
the transverse correlation function as described in \S~\ref{sec:transverse}
and show the result in Figure~\ref{fig:w_dz_data}. At all sightline separations, 
the amplitude for the lower-redshift \cIV systems is larger than that for the 
higher-redshift systems. The magnitude of the increase in clustering strength
is consistent with that measured along the line-of-sight. We estimate
$\xi(2.46) / \xi(3.33) \approx 2$ for \cIV systems.

Between the median subsample redshifts of 3.33 and 2.46,
we expect the amplitude of of the correlation to grow by a factor 1.6.
Our measured correlation functions for the subsamples are consistent
with this growth rate, although the large error bars allow somewhat faster or
slower growth.  Measuring evolution in the scale of the break $r_1$ may be 
possible in the future.  With the current sample, the uncertainties in the 
fitted $r_1$ are large and consistent with no evolution  in $r_1$ larger 
than a factor of two. The addition of larger separation pairs to our analysis 
is important to allow fitting $r_0$ and $r_1$ without additional priors.

\section{SUMMARY} \label{sec:summary}

We obtained echellete spectra of binary quasars, which
probe the IGM along parallel sightlines. Metal-enriched regions
were identified by \cIV absorption systems. We measured their 
clustering in real space, transverse to the sightlines, and
in redshift space, along individual sightlines. At large separations, 
the clustering of intervening \cIV systems is consistent with 
the spatial distribution of galaxies. (Galaxies, as used here, 
referred to Lyman-break galaxies at $z \sim 2.9$ with 
correlation length $r_0 = 4.0$\hinv comoving Mpc.) We found
metals more smoothly distributed than galaxies, however, 
below a scale that we define as the {\it enriched region}.
This scale may represent the turnaround radii of galactic
winds or the overlap of multiple bubbles. We showed that fast outflows would smear out the
\cIV correlation function in redshift space. By comparing
the correlation functions in redshift space and real space,
we found the maximum root-mean-square outflow velocity. We argued 
that the large size of the enriched regions and slow outflow 
speeds are incompatible with dispersal by winds from 
$z \simeq$ Lyman-break galaxies.

We list our empirical results here before discussing their
implications.

\begin{itemize}

\item
We measured redshifts and equivalent widths for 316 intervening \cIV 
systems at $ 1.7 < z < 4.3$. The number of systems per unit redshift 
path falls as the strength of the absorption increases. The distribution
of equivalent widths is consistent with previous measurements of
the column density distribution (Ellison \et 2000; Songaila 2001; 
and Scannapieco \et 2006a).

\item
We counted pairs of \cIV systems as a function of their separation.
The ratio of these counts to the number of pairs in simulated sightlines 
determined the clustering strength (at each separation). The large
clustering amplitude, $\xi > 1$, required special treatment of 
the errors. In an Appendix, we outlined how to compute a correction to 
the Poisson term that characterizes the Landy \& Szalay (1993)
correlation estimator.   We learned that reducing the uncertainty
at fixed separation will require more systems per sightline, i.e., 
more sensitive observations rather than more quasars.

\item
We examined the correlation of \cIV systems between sightlines.
When a \cIV system is present in sightline A, we found a better
than average chance of finding a \cIV system within 1000\kms in
sightline B. The correlation amplitude grows as the separation
of the sightlines decreases. We presented a projection of the 
\cIV correlation function over transverse scales from 
$0.075 \le\ r_p (\hinv {\rm ~comoving~Mpc}) \le 2.84$. 
The fitted break scale, which defines the radius of an
enriched region, was found to be sensitive to the correlation
length. Taking the prior $r_0 \equiv 4.0$\hinv comoving Mpc, 
we found $r_1 = 0.42 \pm 0.15$\hinv comoving Mpc.  However, 
larger enriched regions and correlation lengths improved
the fit statistic slightly, e.g., $r_1 = 0.74 \pm 0.21$\hinv comoving Mpc 
and $r_0 = 4.9$\hinv comoving Mpc. Repeating our analysis 
with more widely separated sightlines will break this degeneracy.

\item
The traditional method for estimating the coherence length
of enriched regions breaks down due to galaxy-galaxy clustering.
We demonstrated that the result cannot be trusted when
the most likely diameter equals the separation of the widest
binary presenting \cIV systems at similar redshift. We propose a revised
maximum-likelihood analysis. When applied to our data,
it converges to a coherence length of 0.3\hinv
comoving Mpc. This result is consistent with our interpretation
of the transverse correlation function.

\item
We measured the clustering of \cIV systems along single sightlines.
We confirmed that the correlation function flattens at small
velocity separation as suggested previously (Rauch \et 
1996; Pichon \et 2003; Scannapieco \et 2006a). Neglecting
peculiar velocities, we fitted a correlation length, $r_0 = 
4.9 \pm 0.7$\hinv comoving Mpc, and  break, $r_1 = 1.2 \pm 
0.3$\hinv comoving Mpc (where $\gamma \equiv 1.6$). Since the 
correlation length was consistent with that fitted to galaxies,
we adopted $r_0 = 4.0$\hinv comoving Mpc (Adelberger \et 2005a) 
and fit $r_1 = 1.08 \pm 0.17$\hinv comoving Mpc. The break
scale in redshift-space space is  larger than the break 
required by the transverse \cIV correlation function. We attribute 
the difference to random, peculiar velocities that impact the 
line-of-sight correlation function but have no affect on the 
transverse correlation function. 

\item
We parameterized the line-of-sight velocity distribution for 
the enriched regions and fitted the line-of-sight correlation
function with $r_0 = 4.0$\hinv comoving Mpc $r_1 = 0.42$\hinv 
comoving Mpc. The resulting root-mean-square peculiar velocity was
$a=120$\kms\ with a $3\sigma$ upper limit of 300\kms. Larger
velocities introduced more power at separations $\Delta v \ge 
a$ than we measure. These speeds may entirely reflect the relative 
motions of galaxies. Because the scale $r_1$ is somewhat larger than 
the virial radius, virial motion within galaxies seem less likely to 
dominate. Most importantly, however, we conclude that the typical 
speeds of outflows over the enriched regions are $\sles\  
200$\kms. The maximum outflow speed is even
lower if the correlation length $r_0$ is larger than assumed.

\item
We measure a higher clustering amplitude for \cIV absorbers at a median 
redshift of 2.5 than we do at redshift 3.3. This growth in clustering
amplitude with cosmic time is consistent with the gravitational 
collapse of large structures. The measurement uncertainties allow
on-going enrichment however.

\end{itemize}

Common \cIV systems in spectra of binary quasars
require enriched regions at least 0.42\hinv comoving 
Mpc in radius. This scale for enriched regions is 
larger than impact parameters
at which Steidel \et (2010) find \cIV absorption. We
attribute the larger size to our higher sensitivity 
to weak \cIV systems. Small, but significant,
redshift-space distortions explain why the line-of-sight correlation 
function flattens on a larger scale. If outflows contribute 
significantly to these peculiar velocities, the outflow velocity 
could be as high as 200\kms. At $z \simeq 3$, 
our smallest fitted size and largest fitted velocity give a minimum 
timescale to disperse the observed metals. This timescale exceeds
exceeds the median age of a Lyman-break galaxy.

In our qualitative picture, galactic winds disperse metals out to a
radius where the outflow stalls or falls back towards the galaxy. If
these bubbles of metals do not overlap, then the size of an enriched 
region measures the turn-around radius of the galactic wind. Large
bubbles of metals can be made consistent with lower outflow speeds by
placing the enrichment at a considerably earlier epoch. 
For example, bubbles blown by galactic winds at $z \sgreat\ 6$, 
need only reach a physical radius of $86$~kpc to spread metals over 
0.42\hinv\ comoving Mpc by redshift 3. The expansion of the universe aids the 
dispersal of metals in this picture. 

The correlation length of the \cIV systems only requires the source 
of the metals be highly biased like Lyman-break galaxies 
(Scannapieco 2005; Scannapieco \et 2006a). Lower mass galaxies at
higher redshift are naturally highly clustered (Scannapieco 2005).
The lower halo mass of early galaxies facilitates the escape of metals
(Dekel \& Silk 1986; Martin 1999). Their large number density may
lead to bubble coalescence enlarging the scale of metal-enriched
regions well beyond that of an individual bubble.

If the metals detected in these \cIV systems were not dispersed by 
galaxies at $z \simeq 3$, then how much earlier did dispersal occur?
The sizes of the enriched regions measured here provide a new
reality check for cosmological simulations that address cosmic
chemical evolution (e.g., Wiersma \et 2009; Oppenheimer \& Dav\'{e} 2008; 
Kobayashi \et 2007; Scannapieco \et 2006b).
Some simple arguments, however, suggest $z \sgreat\
4.3$. Consider galaxies at $z=4.3$ with the same bias as Lyman-break
galaxies. The halo mass of such galaxies would be about 0.2 times
that of a typical LBG. The comoving space density of such halos 
is about 5 times higher than that of LBGs. In the 723~Myr between redshift 
4.3 and 3.0, an outflow traveling at 200\kms would reach 0.55\hinv comoving 
Mpc. Some of these galaxies would still be around at redshift 3
and recognizable by their older stellar population and low mass.
(The expansion rate of the wind exceeds the cosmic expansion rate,
so the wind is unlikely to stall.) Taking a cross section of
$\pi r_1^2$, we estimate the density of \cIV systems (down to our
sensitivity of 20\mA) would be $dN/dz \sim 5$. This value is
comparable to our data if there are 2 to 3 \cIV system per galaxy.
We obtain higher $dN/dz$, and therefore more systems per galaxy
if we place enrichment at higher redshift. Only of order 10\% of
of the \cIV population is easily attributed to enrichment at redshift
3 by Lyman-break galaxies.

The results desribed above illustrate how quasar tomography can
be used to map the distribution of intergalactic metals.
The sightlines used in this paper were chosen for the relatively
small transverse separation of the binary quasars. It is possible to 
observe binaries at wider separation. When the correlation of
metal-line systems is measured over a very broader range in scale, 
it will be possible to simultaneously fit $r_1$, $r_0$, and $\gamma$ 
to the transverse correlation function. 
The feasibility of making sensitive observations along many
pairs of sightlines will grow enormously with the next generation
of giant ground-based telescopes. Our results suggest that
measuring the evolution in the size of enriched regions with redshift 
would directly reveal the IGM enrichment history.


\acknowledgements
We thank Nicolas Bouch\'{e}, Jon Oiler, and Rob Thacker for discussions that
improved the content of this paper.
C.L.M. acknowledges support for this work from the David and Lucile Packard Foundation and 
NSF grant AST-080816. E.S. acknowledges support from NASA theory grant NNX09AD10G. 
S.G.D. acknowledges partial support from NSF grants AST-0407448 and 
AST-0909182, and the Ajax Foundation. The authors wish to recognize and acknowledge the very 
significant cultural role and reverence that the summit of Mauna Kea has always had within 
the indigenous Hawaiian community.  We are most fortunate to have the opportunity to 
conduct observations from this mountain. We thank Mr. Kurt Soto, Dr. Grant Hill, 
and Dr. Greg Wirth for assistance during the observing runs.

{\it Facilities:}  \facility{Keck}

\clearpage

\appendix







\section{Appendix: Computing The Variance of Our Correlation Measurements}

\subsection{Single Sightlines}

First we consider measurements along single lines of sight. Numbering all the sightlines in the sample with an index $\ell$, we can compute
the variance in the correlation function for a bin $k$, representing separations from $v-\Delta v/2$ to $v+ \Delta v/2$ as
\be
\sigma^2 (\xi_k)  = \frac{1}{\sum_l \left< RR \right>^\ell_k} \times \sum_l \sigma^2(\left<DD\right>_k^\ell),
\label{sigk}
\ee
where $\left<D D \right>_k^\ell$ is the measured number of pairs  in sightline $\ell$ and bin $k$, 
$\left< RR\right>_{k}^\ell$ is the expected number of pairs in sightline $\ell$ 
and bin $k$ for a random distribution, such that
 $\sum_\ell \left< RR\right>_{k}^\ell$  is the total number of expected pairs
in bin $k$ in the full data set, and $\sigma^2(\left<D D \right>_k^\ell)$ is the variance in $\left<D D \right>_k^\ell.$

From Poisson statistics $\sigma^2(\left<D D \right>^\ell_k) = \left<D D \right>_k^\ell$, however as discussed by 
Mo \etal (1992), this is an underestimate of the full variance.
To compute the full variance of a single sightline we  divide it into many small 
redshift bins with an index $i$. For simplicity we drop the  superscript $\ell$ and
let $N_i$ be the number of absorbers in one of these bins, 
where the bins are small so that $N_i$ is always zero or 1.
To compute the number of pairs in a given sightline, one must compute 
$\left<DD \right>_k = \frac{1}{2}\sum_{i,j} N_i N_j U_{i,j}$ 
where $U_{i,j}$ is 1 if the velocity difference between line of sight bins
$i$ and $j$ is greater than $v-\Delta v/2$ and less than $v+\Delta v/2,$ and $U_{i,j}=0$ otherwise.

If we square this we can get an estimate of the variance
\be
\frac{1}{4} \sum_{i,j,m,n} N_i N_j N_m N_n U_{i,j} U_{m,n},
\ee
Here  there are three possibilities, one is that $i$, $j,$ $m,$ and $n$ are all different,
one is that either $i$ or $j$ falls in the same bin as either $m$ or $n$, and one is that both $i$ and $j$ are matched
with $m$ and $n$. This gives us three terms.  If all  the bins are different, this gives the average correlation function squared, which does not contribute to the variance. If there are two matches between bins, this  gives a term that goes as the number of pairs, which leads to the Poisson term.  Finally, if there is only one match, we obtain a term of the form
\be
\sum _{i,j,m} \left<N_i N_j N_m \right> U_{i,j} U_{i,m}.
\label{eq:3term}
\ee
There is no factor of 1/4 here because there are 4 possible pairings that contribute to this term.
The term is equal to the joint probability of finding three absorbers: two of them  at
a distance $v \pm \Delta v/2$ from the first one.   

There are two possibilities for this distribution. One in which 
absorption-line systems
$j$ and $m$ are on opposite sides of absorber $i. $  This possibility contributes 
\be
N_{\rm abs} \times \left\{ 2 n \Delta v [1+\xi(v)] \right\}  \times \left\{ n \Delta v [1+\xi(v)] \right\}, 
\ee
where $N_{\rm abs}$ is the total number of absorption-line systems in a given sightline and $n$ is the average number of absorption-line systems per unit length.  There is a two on the second term because absorber $j$ can either lie at higher or lower redshift than absorber $i$, but this factor does not appear in the third term because we are assuming that $j$ and $m$ are on opposite sides.  Note that we are also ignoring the 3-point correlation function,  the (usually small) excess probability of any absorber being located in bin $k$ above due to {\em both} $i$ and $j$.

The second possibility for eq.\ (\ref{eq:3term}) is  that  absorption-line systems $j$ and $m$ are on the same side of absorber $i. $  In this case it is more important that $j$ and $m$ are near each other than at a distance $v$ from absorber $i$.  So this contribution looks like
\be
N_{\rm abs} \times \left\{ 2 n \Delta v [1+\xi(v)] \right\}  \times \left\{ n \Delta v [1+\bar {\bar \xi}(v)] \right\}, 
\ee
where we define
\be
\bar {\bar \xi} (\Delta v) = \left[ \Delta v^{-2}  \int_0^{\Delta v}  dv'  \int_0^{\Delta v}  dv'' \xi(v'-v'') \right],
\ee
where the double over bar indicates an average over {\em two} positions,
all positions in which absorption-line systems $j$ and $m$ could lie within a bin of size $\Delta v.$
Again, we have ignored the contribution from the 3 point function, now considering only the excess probability due to the fact that $j$ and $m$ are in the same bin.

Summing up these two contributions and factoring out $\left<D D \right>_k^\ell$ where
\be
\left<D D \right>_k = \frac{N_{\rm abs}}{2} \left\{ 2 n \Delta v [1+\xi(v)] \right\}, 
\ee
gives
\be
\sum_{i,j,m} \left<N_i N_j N_m \right> U_{i,j} U_{i,m}
=
2 \left<D D \right>_k   2 n \Delta v
\left[ 1+\frac{\xi(v) + \bar {\bar \xi}(\Delta v)}{2}  \right]\\ \nonumber
\ee
such that the total variance in sightline $\ell$ is given by 
\be
\sigma^2(\left<D D \right>_k^\ell) =  \left<D D \right>_k^\ell + 
\frac{4 (\left<D D \right>_k^\ell)^2}{N_{\rm abs}}
 \left\{  1+ \frac{ \bar{\bar \xi} (\Delta v) -\xi(v)  }{2[1+\xi(v)]} \right \}.
 \label{eq:standarvariance} 
\ee
In the case in which the correlation function is small, the term in the brackets is just 1, recovering the extra contribution to the variance as computed in  Mo \etal (1992); but if we know $\xi(v)$, then we can also evaluate the term in brackets.

This variance can be reduced by working with the Landy and Szalay estimator for the correlation function, which for a single
sightline is given by $\frac{DD-2DR+RR}{RR}.$    To compute the variance of this estimator we have to consider two additional terms.  The
first of these arises from the cross-correlation between $DD$ and $DR$ takes the form
\be
- 2\sum_{i,j,k,m} N_i N_j N_m R_n U_{i,j} U_{m,n},
\ee
where there is a 2 because only the $\left<DD \right>$ term is defined with a $1/2$ in front.
Again we have a term in which $i$, $j$, $m$, and $n$ are all different which does not contribute to the variance, and a term in which $i$ and $m$ or $j$ and $m$ are paired, which gives
\be
-4 \sum_{i,j,n} \left<N_i N_j R_n \right> U_{i,j} U_{i,n}.
\ee
When appropriately normalized to account for the number of absorption-line systems in the random data set
this
\be
-2  N_{\rm abs} [2n \Delta v (1+ \xi(v) ] 2 n \Delta v =  -4 \left<D D \right>_k 2 n \Delta v.
\ee
Finally from the term 4$\left<DRDR \right>$ we have a term in which $N_i$ and $N_k$ are pared which, when normalized, gives
\be
N_{\rm abs} [2n \Delta v] ^2 = 2 \left<D D \right>_k \frac{2 n \Delta v}{1 + \xi(v)}.
\ee
Putting this together we get that for the Landy and Szalay estimator, for a single sightline, the total variance is equal to the Poisson term, plus a second 
a term
\be
2 \left<D D \right>_k 2 n \Delta v
\left[ -1+ \frac{1}{1+\xi(v)}  + \frac{\xi(v) + \bar {\bar \xi}(\Delta v)}{2} \right],
\ee
such that the variance in sightline $\ell$ is given by 
\be
\sigma^2(\left<D D \right >_k^\ell) =  \left<D D \right>_k^\ell + 
\frac{4 (\left<D D \right>_k^\ell)^2}{N_{\rm abs}}
\left[ \left(\frac{\xi(v)}{1+\xi(v)} \right)^2  +   \frac{\bar {\bar \xi} (\Delta v) -\xi(v)  }{2(1+\xi(v))} \right].
\label{eq:lzvariance}
\ee

We do not know the correlation function to evaluate equations  (\ref{eq:standarvariance}) and (\ref{eq:lzvariance}), but as an
upper limit on the variance we can take the galaxy-galaxy correlation function, taking into account the resolution of our data.
This is a power law with a slope $\gamma$ and correlation length $v_0 \equiv H(z) r_0,$ above $v_{\rm res} = 50$ km/s, and 0 below that.
That is 
\be
\xi( v) =  (v/v_0)^{-\gamma},
\ee
if  $v \ge v_{\rm res}$ and $\xi(v) = 0$ otherwise,
where $r_0$ and $\gamma$ are taken to fit the galaxy-galaxy correlation function.
In this case
\be
\bar {\bar \xi}(\Delta v) = 
 \left(\frac{\Delta v}{v_0}\right)^{-\gamma} \frac{2}{(2-\gamma)(1-\gamma)}+
\left(\frac{v_{\rm res}}{v_0}\right)^{-\gamma} \left[ \left(\frac{v_{\rm res}}{\Delta v} \right)^2  \frac{2}{(2-\gamma)} - \frac{v_{\rm res}}{\Delta v} \frac{2}{1-\gamma} \right],
\ee
 if  $\Delta v \ge v_{\rm res}$ and $\bar{\bar \xi}{\Delta v} = 0$ otherwise.

\subsection{Pairs of Sightlines}

Moving on to the transverse correlation function between two sightlines $A$ and $B$ that make up a pair $\ell,$ the simplest estimator is
\be
1+ \xi^T_{k} = \frac{1}{\sum_{\ell \in k}  \left< R^A R^B  \right> } \sum_{\ell \in k} \left < D^A D^B \right>^\ell,
\ee
where the sum is over all pairs $\ell$ whose separations put them in a traverse bin $k$.
The case we are interested in is the one in which $\left<D^A D^B \right>^\ell$ includes all absorption-line systems that are separated
by 0 km/s up to some maximum value,  $\Delta v.$ 

Let us again consider the variance for a single sightline.
Following our notation from the previous section,
we define $N_{\rm abs}^A$ and $N_{\rm abs}^B$ as the number of absorption-line systems in sightline $A$ and $B$ in this pair, and $n^A$ and $n^B$ as the
number density of absorption-line systems in sightline $A$ and $B.$
In this case
\be
\left<D^A D^B \right>^\ell = N^A_{\rm abs} \,  2 \Delta v   n^B \, [1+\bar \xi(r_p,\Delta)] =  N^B_{\rm abs} \,   2 \Delta v n^A \, [1+\bar \xi(r_p,\Delta v)],
\ee
where $\bar \xi(r_p,\Delta v) \equiv \frac{1}{\Delta v} \int_0^{\Delta v} dv'  \xi(r_p,v'),$ such that the single bar denotes a single velocity 
average of  $\xi(r_p,v')$ the correlation function between two absorption-line systems at a transverse comoving distance of $r_p$ and comoving separation of $v'$.

To compute the variance of a single pair of sightlines, we again divide both sightlines into many small bins and square this gives us
\be
\sum_{i,j,m,n} N_i^A N_j^B N_m^A N_n^B U_{i,j} U_{m,n}.
\ee
This leads to a term where $i$, $j$, $m$ and $n$ are all different, which gives the mean value squared, a term where $i=m$ and $j=n,$ which 
corresponds to the Poisson term, and a new term
\be
\sum_{i,m,n} \left<N_i^A N_m^A N_n^B \right> U_{i,n} U_{m,n} + \sum_{i,j,n} \left<N_i^A N_j^B N_n^B \right> U_{i,m} U_{j,m},
\label{eq:AB}
\ee

Focusing on the first of the two sums in eq.\ (\ref{eq:AB}),  
we can simplify this by considering only the case in which $i$ is at a lower redshift than $n$, and then multiplying by 2. This gives
\be
\sum_{i,n,m} \left<N_i^A N_m^A N_n^B \right> U_{i,m} U_{n,m}
= N^B_{\rm abs} \left\{2 n^A \Delta v  \left[1 + \bar \xi(r_p,\Delta v) \right] \right\} \left\{2 n^A \Delta v \left[ 1+ \bar {\bar {\bar \xi}}(r_p,\Delta v) \right] \right\}
\ee
where
\be
\bar {\bar {\bar \xi}}(r_p,\Delta v) \equiv 
\frac{1}{[1+ {\bar \xi(r_p,\Delta v)] 2 (\Delta v)^2 }}
 \int_0^{\Delta v} d v' [1+\xi(r_p,v')] \int_{-\Delta v}^{\Delta v} dv'' \left\{ \max \left[\xi(r_p,v''),\xi(0,|v''-v'|) \right ] \right\} \nonumber
\ee
Factoring out $ \left<D^A D^B \right>^\ell$ this gives
\be
\sum_{i,n,m} \left<N_i^A N_n^A N_m^B\right> U_{i,m} U_{n,m}
= \left<D^A D^B \right>^\ell  2 n^A \Delta v  \left[1+\bar {\bar {\bar \xi}}(r_p,\Delta v) \right].
\ee
Computing the second term in eq.\  (\ref{eq:AB})  in a similar manner we get
\be
\sum_{i,j,m} \left<N_i^A N_j^B N_m^B \right> U_{i,j} U_{i,m}
= \left<D^A D^B \right>^\ell  2 n^B \Delta v \left[1+\bar {\bar {\bar \xi}}(r_p,\Delta v) \right].
\ee
Thus the variance for pair $\ell$ is
\be
\sigma^2(\left<D^A D^B \right>^\ell) =  \left<D^A D^B \right>^\ell + 
\left(\left<D^A D^B \right>^\ell \right)^2 \left[\frac{1}{N_A}+\frac{1}{N_B}\right]
 \left[  1+ \frac{\bar {\bar {\bar \xi}} (r_p,\Delta v) -\bar \xi(r_p,\Delta v)  }{1+\bar \xi(r_p, \Delta v)} \right].
 \label{eq:ABstandarvariance} 
\ee

In the case of the Landy and Szalay estimator, which now goes as $\frac{D^AD^B - D^A R^B - R^A D^B + R^A R^B}{R^A R^B},$
two new cross terms arise.  The first is a term of the form $-2 \left( \left<D^AD^BD^A R^B \right> +  \left<D^AD^B R^A D^B \right> \right)$ which leads to
\be
-2 \left( \sum_{i,j,m} N_i^A N_j^B R_m^A  U_{i,j} U_{j,m}  + \sum_{i,j,n} N_i^A R_j^B N_n^B U_{i,j} U_{j,n} \right) = -
2 \left<D^AD^B \right>^\ell 2 \Delta v (n^A+ n^B).
\ee
The second is a term $\left<D^A R^B D^A R^B \right> + \left<R ^A D^B R^A D^B \right>,$ which leads to
\be
\left( \sum_{i,j,m} R_i^A R_m^A N_j^B U_{i,j} U_{j,m}  + \sum_{i,j,m,n} N_i^A R_m^B N_j^B U_{i,j} U_{j,m} \right)
= \left<D^AD^B  \right>^\ell 2 \Delta v \frac{(n^A+ n^B)}{1 + \bar \xi(r_p,\Delta v)}.
\ee
Summing these together then gives 
\be
\sigma^2 \left(\left<D^A D^B \right >^\ell \right) =  \left<D^A D^B \right>^\ell + 
\left(\left<D^A D^B \right>^\ell \right)^2 \left[\frac{1}{N_A}+\frac{1}{N_B}\right]
\left[ \left(\frac{\bar \xi(r_p,\Delta v)}{1+\bar \xi(r_p,\Delta v)} \right)^2  +   \frac{\bar{\bar {\bar \xi}} (r_p,\Delta v) -\bar \xi(r_p,\Delta v)  }{1+\bar \xi(r_p,\Delta v)} \right].
\label{eq:ABlzvariance}
\ee
In the case of our data, we can estimate $\bar \xi(r_p,\Delta v)$ and $\bar {\bar {\bar \xi}}(r_p,\Delta v)$ by carrying out appropriate integrals over
\be
\xi(r_p,v) = 
\left[ \frac{r_p^2 + \Pi ^2}{r_0^2} \right]^{-\gamma/2} 
\ee
if  $v \ge v_{\rm res}$ or $r_p \ne 0$ and $\xi(r_p,v)=0$ otherwise,
where $\Pi = v/H(z)$ and $r_0$ and $\gamma$ are taken to fit the galaxy-galaxy correlation function.



\clearpage

\references

Adelberger, K. L. \& Steidel, C. C., Shapley, A. E., \& Pettini, M.  2003, \apj, 584, 45


Adelberger, K. L. \et 2005a, \apj, 619, 697 

Adelberger, K. L. \et 2005b, \apj, 629, 636 

Aguirre, A. \et 2001, \apj, 560, 599

Becker, G., Rauch, M. \& Sargent, W.L.W.  2009, \apj, 698, 1010 


Cooksey, K. L. \et, \apj, 708, 868  

Cooray, A. \& Sheth, R. 2002, PhR, 372, 1

Coppolani, F. \et 2006, \mn, 370, 1804 

Crotts, A. P. S. \et 1994, \apj, 437, L79

Davis, M. \& Peebles, J. E. 1983, \apj, 267, 465 

Dekel, A \& Silk, J. 1986, \apj, 303, 39

Dekel, A. \& Birnboim, Y. 2008, \mn, 383, 119


Dinshaw N., Weymann R. J., Impey C. D., Foltz C. B., Morris S. L., Ake T., 1997, ApJ, 491, 45 

D'Odorico, V. \et 2010, \mn, 401, 2715


Ellison, S., Songaila, A., Schaye, J, \& Pettini, M \et 2000, \aj, 120, 1175 

Ellison, S. L. \et 2004, \aa, 414, 79 



Ferrara, A. Pettini, M. \& Shchekinov, Y. 2000, \mn, 319, 539 


Furlanetto, S. R. \& Loeb, A. 2003, \apj, 588, 18 


Gnedin \& Ostriker 1997, ApJ, 486, 581 


Hawkins, E. \et 2003, \mn, 346, 78

Hennawi, J. F., 2004, PhD Dissertation, Princeton University, Publication Number: AAT 3151085 

Hennawi, J. F. \et , 2006a, \aj, 131, 1 

 Hennawi, J. F., et al.\ 2006b, \apj, 651, 61 

Hennawi, J. F. \et 2010, arXiv:0908.3907v1, submitted to ApJ

Kawata, D. \& Rauch, M. 2007, \apj, 663, 38 

Kobayashi, C. \et 2007, \mn, 376, 1465


Kriss, G. 1994 PASP  Conf.  Series, Vol.  61,  p.437 

Landy, S. D. \& Szalay, A. S. 1993, \apj, 412, 64 

Lopez, S.  \et 2000, \aa, 357, 37

Lu, L.  1991, \apj, 379, 99

Madau, P., Ferrara, A. \& Rees, M. J. 2001, \apj, 555, 92 

Martin, C. L. 1999, \apj, 513, 156

Martin, C. L., Kobulnicky, H. A., \& Heckman, T. M. 2002, \apj, 574, 663

Martin, C. L. 2005, \apj, 621, 227

McGill C., 1990, MNRAS, 242, 544 

Meyer, D. M. \& York, D. G. 1987, \apj, 315, 5 

Mo, H. J., Jing, Y. P., \& B\"{o}rner, G. 1992, \apj, 392, 452 

Morton, D. 2003, \apjs, 149, 205

Oppenheimer, B. D., Dav\'{e}, R., 2008, \mn, 387, 577 

Oppenheimer, B. D., Dav\'{e}, R. \& Finlator, K.  2009, \mn, 396, 729 

Petitjean, P. \& Bergeron, J. 1990, \aa, 231, 309 

Pettini, M. \et 2002, \apj, 569, 742 


Pichon, C. \et 2003, \apj, 597, L97   

Qian, Y.-Z.,  Sargent, W. L. W. \& Wasserburg, G. J. 2002, ApJ, 569, 61 

Rauch, M., Sargent, W. L. W., Womble, D. S., Barlow, T. A. 1996, \apj, 467, 5 

Rauch, M. Sargent, W. L. W. \& Barlow, T. A. 1999 \apj, 515, 500 

Rauch, M., Sargent, W. L. W., \& Barlow, T. A. 2001, \apj, 554, 823 


Rauch, M. \et 2002, \apj, 576, 45


Rupke, D. \et 2005, \apjs, 160, 115 

Ryan-Weber, E. V. 2009, \mn, 395, 1476 

Sawicki, M. \& Yee, H. C. 1998, \aj, 115, 1329

Scannapieco, E., Ferrara, A., \& Madau, P. 2002, \apj, 574, 590 

Scannapieco, E. 2005, \apj, 624, 1

Scannapieco, E. \et 2006a, \mn, 365, 615 

Scannapieco, E. \et 2006b, \mn, 371, 1125 

Schaye, Y. \et 2003, \apj, 596, 768



Shapley, A. E. \et 2001, \apj, 562, 95 

Shapley, A. E. \et 2003, \apj, 588, 65 

Sheinis, A. I \et 2002, \pasp, 114, 851 

Songaila, A. \& Cowie, Lennox L. 1996, \aj, 112, 335 

Songaila, A. 2001, \apj, 561, L154 

Steidel, C. \et 2010, arXiv:1003.0679

Theuns, T. \et 2002, \apj, 578, L5 


Tytler, D. \et 2009, \mn, 392, 1539  


Weiner, B. J. \et 2009, \apj, 692, 187

Wiersma, R. P. C. \et 2009, \mn, 399, 574




\begin{deluxetable}{llllllll}
\tablecaption{ESI Spectroscopy}
\tablehead{
\colhead{Paired Quasars}	    &	
\colhead{ $z_{QSO}$}   & 
\colhead{r}       & 
\colhead{$\theta$}  & 
\colhead{SNR} & 
\colhead{Date Observed} &
\colhead{$\tau$}  
\\
\colhead{ }    & 
\colhead{Redshift}   & 
\colhead{(AB)}& 
\colhead{(\arcsec)} & 
\colhead{} &
\colhead{} &
\colhead{(s)}    
\\
\colhead{ (1) }    & 
\colhead{ (2)}   & 
\colhead{(3)}& 
\colhead{(4)} & 
\colhead{(5)} &
\colhead{(6)} &
\colhead{(7)}    
}
\startdata
SDSSJ0117+3153B  &   2.640     & 19.86   & 11.3    & 29.0 & 2005 Nov. &  17280  \\
SDSSJ0117+3153A  &   2.625     & 20.52   &	''      & 18.5 & 2005 Nov. &  17280 \\	 
SDSSJ0225+0048C &2.700 & 20.40  & 100.3, 75.3 & 12.8 & 2006 Nov., 2008 Jan. &  6771    \\
SDSSJ0225+0048A &2.820 & 20.54 & 100.3, 27.4  & 12.6 & 2005 Nov. &  7100 \\
SDSSJ0225+0048B &2.757 & 20.40 & 75.3, 27.4   & 16.5 & 2008 Jan. &  7260  \\
SDSSJ0245-0113A  &	2.462     & 19.63   & 4.5	& 20.2 & 2006 Nov. &  12400 \\
SDSSJ0245-0113B  &	2.459	  & 20.51   & ''	& 16.5 & 2006 Nov. &  12400 \\
SDSSJ0818+0719A   &4.618 & 19.66 & 132.9      & 19.8 & 2006 Nov. &  1800   \\
SDSSJ0818+0718B   &4.178 & 20.08 & ``         & 19.8 & 2008 Jan. &  1800  \\ 
SDSSJ0925+3859A  &	3.135     & 19.32   & 57.1	& 22.5 & 2005 Nov. &  6105 \\
SDSSJ0925+3859B  & 2.823	  & 20.11   & ''	& 20.8 & 2005 Nov. &  10800 \\
SDSSJ0956+2643A  &	3.083     & 19.28   & 16.5	& 19.0 & 2007 Apr. &  2400  \\
SDSSJ0956+2643B  & 3.083	  & 20.52   & ''	& 14.8 & 2007 Apr. &  14740 \\
SDSSJ0959+1033A   &4.021 & 19.34 & 44.1       & 33.8 & 2005 Nov. &  12500  \\ 
SDSSJ0959+1033B   &4.021 & 20.11 & ``         & 11.2 & 2006 Nov. &  6300  \\ 
SDSSJ1021+1112A &	3.830	  & 20.60   & 7.6	& 45.7 & 2007 Apr. &  21600 \\ 
SDSSJ1021+1112B &  3.830	  & 20.71   & ''	& 24.8 & 2007 Apr. &  21600 \\
SDSSJ1026+4614A   &3.421 & 19.73 & 37.1       & 24.0 & 2008 Jan. &  3600   \\ 
SDSSJ1026+4614B   &3.345 & 19.73 & ``         & 17.0 & 2008 Jan. &  3600  \\ 
SDSSJ1116+4118A & 	3.000	  & 18.18   & 13.8	& 33.3 & 2006 Mar. &  3291 \\
SDSSJ1116+4118B & 	3.000	  & 19.19   & ''	& 18.0 & 2006 Mar.  &  3291 \\
SDSSJ1144+0959A &  3.146	  & 18.33   & 23.5	& 33.6 & 2007 Apr. &  2400 \\
SDSSJ1144+0959B &  2.974	  & 19.87   &		& 23.1 & 2007 Apr. &  10800 \\
SDSSJ1228+0232A   &3.138 & 20.42 & 47.3       & 11.7 & 2008 Jan. &  3600   \\ 
SDSSJ1228+0232B   &3.095 & 19.83 & ``         & 11.3 & 2008 Jun. &  2400  \\ 
SDSSJ1248+1957A   &3.876 & 20.23 & 64.8       & 14.6 & 2008 Mar.,2008 June &  4800  \\ 
SDSSJ1248+1957B   &3.817 & 19.55 & ``         & 16.5 & 2008 Mar.,2008 June &  3000  \\ 
SDSSJ1307+0422A  &	3.024	  & 18.05   & 8.2	& 55.3 & 2007 Apr. &  3600 \\
SDSSJ1307+0422B  & 3.024	  & 19.33   & ''        & 22.3 & 2007 Apr.  &  3600 \\  
SDSSJ1314+2818A  &4.821 & 21.27 & 116.6      & 18.6 & 2008 July &  2400  \\   
SDSSJ1314+2818B  &4.390 & 20.92 & ``         & 7.5 & 2008 July &  900  \\   
SDSSJ1353+4852A   &3.848 & 20.43 & 37.1       & 10.1 & 2008 July &  2400  \\   
SDSSJ1353+4852B   &3.848 & 20.91 & ``         & 15.2 & 2008 July &  3600  \\   
SDSSJ1420+2830A   &4.312 & 20.33 & 10.9       & 20.8 & 2008 Jun. &  10800   \\   
SDSSJ1420+2830B   &4.290 & 20.77 & ``         & 20.5 & 2008 Jun. &  10800  \\   
SDSSJ1513-0131A & 3.224	  & 19.23   & 48.1	& 22.0 & 2007 Apr. &  4200 \\
SDSSJ1513-0131B & 3.191	  & 19.91   & ''        & 18.1 & 2007 Apr.  &  7200 \\
SDSSJ1627+4606A &	4.110	  & 20.66   & 34.1	& 20.4  & 2007 Apr. &  10800 \\
SDSSJ1627+4606B & 3.812	  & 20.15   & ''	& 23.1 & 2007 Apr.  &  9600\\
SDSSJ1054+0216A &  3.976     & 19.53   & 88.2      & 16.2 & 2007 Apr. &  2400 \\
SDSSJ1054+0216B &  3.976     & 18.81   & ''        & 19.8 & 2007 Apr. &  2400 \\
SDSSJ1215-0309A&   4.001     & 20.20   & 47.1     & 13.3 & 2007 Apr. &  2400 \\
SDSSJ1215-0309B&   3.672     & 19.81   & ''       & 15.4 & 2007 Apr. &  2400 \\
SDSSJ1404+4005A &	4.027	  & 19.62   & 47.3 	& 17.6 & 2007 Apr. &  2556 \\
SDSSJ1404+4005B & 	4.027	  & 21.01   & ''        & 13.3 & 2007 Apr.  &  7200 \\
SDSSJ1541+2702A   &3.621 & 20.62 & 6.4        & 14.1 & 2008 Jan., 2008 Mar. &  13782   \\   
SDSSJ1541+2702B   &3.626 & 20.71 & ``         & 10.4 & 2008 Jan., 2008 Mar. &  13782   \\   
SDSSJ1542+1733A   &3.261 & 20.06 & 27.4       & 15.2 & 2008 June &  3000    \\   
SDSSJ1542+1733B   &2.782 & 18.87 & ``         & 18.5 & 2008 June &  1200   \\
SDSSJ1622+0702A   &3.256 & 20.33 & 5.8        & 14.9 & 2008 July &  3000 \\
SDSSJ1622+0702B   &3.260 & 17.19 & ``         & 86.5 & 2008 July &  3000 \\
SDSSJ1627+2215B   &3.251 & 20.55 & 35.6       & 11.4 & 2008 June &  5400 \\
SDSSJ1627+2215A   &3.709 & 19.29 & ``         & 12.7 & 2008 June &  3000  \\ 
SDSSJ2157+0015A   &2.560 & 20.55 & 20.8       & 12.7 & 2006 Nov. &  4257 \\   
SDSSJ2157+0015B   &2.542 & 19.39 & ``         & 11.6 & 2006 Nov. &  3000 \\ 
\hline
\enddata
\tablecomments{
(1) Quasar name.
(2) Quasar redshift. As decribed in Hennawi \et (2009),
we measure the wavelegnths of \cIV, \ion{C}{3}, and \ion{Si}{4} and apply
the typical shift found between these high-ionization lines and
the systemic frame, as traced by the Mg~II doublet (Shen \et 2007).
(3) Quasar r-band magnitude.
(4) Angular separation of pair. Two values indicate the triplet system.
(5) Median continuum S/N ratio in the bandpass between \lya\ and \cIV;  the continuum
within 5000\kms of either line is excluded from the distribution.
(6) Date Observed.
(7) Total exposure time.
}
\label{tab:esiALL} 
\end{deluxetable}

\end{document}